\documentclass[letterpaper,hyper]{JHEP3}
\usepackage{epsfig}
\usepackage{amsmath, amsfonts,euscript,bbm}
\usepackage{ifthen}
\usepackage{graphicx}

\newcommand{\be}{\begin{eqnarray}}

\newcommand{\ee}{\end{eqnarray}}

\newcommand{\Mpl}{M_{p}}

\newcommand{\ex}[1]{\langle #1 \rangle}

\newcommand{\R}{\mathcal{R}}

\newcommand{\rz}{\bar \rho}

\newcommand{\uu}{u_{i}}
\newcommand{\ut}{u^{(2)}_{i}}
\newcommand{\ru}{\delta}
\newcommand{\rt}{\delta^{(2)}}

\newcommand{\no}{\delta n}
\newcommand{\nbod}{\dot{\delta n}_{b}}

\newcommand{\nt}{\delta n^{(2)}}
\newcommand{\mmu}{\chi}
\newcommand{\tzr}{\Theta_{0,r}}
\newcommand{\tru}{\Theta_{(1),r}}
\newcommand{\trt}{\Theta_{(2),r}}
\newcommand{\tcu}{\Theta_{(1)}}
\newcommand{\tct}{\Theta_{(2),0}}

\newcommand{\rgz}{\bar{\rho}_{\gamma}}
\newcommand{\rgo}{\delta_{\gamma} }
\newcommand{\rgt}{\delta_{\gamma(2)} }
\newcommand{\rgod}{\dot \delta_{\gamma}}
\newcommand{\rgtd}{\dot \delta_{\gamma(2)}}
\newcommand{\rgodd}{\ddot \delta_{\gamma}}
\newcommand{\rgzd}{ \dot{\bar \rho}_{\gamma}}
\newcommand{\rbz}{\bar{\rho}_{b}}
\newcommand{\rbo}{\delta_{b}}
\newcommand{\rbt}{\delta_{b(2)}}
\newcommand{\rbod}{ \dot{\delta}_{b}}
\newcommand{\rbtd}{ \dot{\delta}_{b(2)}}
\newcommand{\nbz}{ {\bar n}_{b}}
\newcommand{\nbt}{  \delta n_{b(2)}}
\newcommand{\nbzd}{ \dot{\bar n}_{b}}
\newcommand{\ngzd}{ \dot{\bar n}_{\gamma}}

\newcommand{\ngt}{ {\delta n}_{\gamma(2 )}}
\newcommand{\ngo}{ {\delta n}_{\gamma}}
\newcommand{\ngz}{ {\bar n}_{\gamma}}
\newcommand{\ngg}{ n_{\gamma}}
\newcommand{\Tz}{\bar T}
\newcommand{\Tzd}{\dot{\bar T}}
\newcommand{\nsq}{\partial_{i}\partial^{i}}
\newcommand{\dm}{\delta \mu}
\newcommand{\dmd}{\dot {\delta \mu}}
\newcommand{\dto}{\Theta_{r}}
\newcommand{\dtt}{\Theta_{r,(2)}}
\newcommand{\qmd}{q_{\mu D}}
\newcommand{\mth}{ \mu_{\rm th}}
\newcommand{\mthb}{ \mu_{b,\rm th}}
\newcommand{\Ao}{A_{\rho}}
\newcommand{\At}{A_{n}}
\newcommand{\Atr}{A_{s}}
\newcommand{\ar}{a_{R}}
\newcommand{\br}{b_{R}}

\title{A Hydrodynamical Approach to CMB $\mu$-distortions}

\author{Enrico Pajer${}^{1}$ and Matias Zaldarriaga${}^{2}$, \\
$^1$ Department of Physics, Princeton University, Princeton, NJ 08544, USA\\
$^2$ Institute for Advanced Study, Princeton, NJ 08544, USA}

\date{}
\abstract {Spectral distortion of the cosmic microwave background provides a unique opportunity to probe primordial perturbations on very small scales by performing large-scale measurements. We discuss in a systematic and pedagogic way all the relevant physical phenomena involved in the production and evolution of the $\mu$-type spectral distortion. Our main results agree with previous estimates (in particular we show that a recently found factor of $3/4$ arises from relativistic corrections to the wave energy). We also discuss several subleading corrections such as adiabatic cooling and the effects of bulk viscosity, baryon loading and photon heat conduction. Finally we calculate the transfer function for $\mu$-distortions between the end of the $\mu$-era and now.  }
\preprint{arXiv:1206.xxxx}
\begin{document}

\tableofcontents


\begin{table}[!h]
\centering
\begin{tabular}{|c|l|l|c|}  \hline
Symbol    & Relation &	Meaning 	&  Equation\\ \hline\hline
$a$	  & 	& scale factor 	&	\eqref{FLRW}	\\ \hline
$\tau$ & $a d\tau=dt$ & conformal time & \\ \hline
$H$ &$=\dot a/a$ & Hubble parameter & \\ \hline
$\Mpl$ & $= \left(8\pi G_{N}\right)^{-1/2}$ & reduced Planck mass & \\ \hline
$R$ & $=3\rgz/(4\rbz)$ & baryon loading & \\ \hline
$R_{\nu}$ & $=\bar{\rho_{\nu}}\rgz$ & neutrino-to-photon ratio & \eqref{Rnu} \\ \hline
$\rho_{x}$ & $=u_{\mu}u_{\nu} T^{\mu\nu}_{x}$ & energy density of the specie ``$x$'' & \\ \hline
$\rgz$ & $=\bar T^{4} \,\pi^{2}/15$ & background photon energy density & \eqref{ex3} \\ \hline
$p_{x}$ &  & pressure of the specie ``$x$'' & \\ \hline
$n_{x}$ & $=-u_{\mu}N^{\mu}_{x}$ & number density of specie ``$x$'' & \\ \hline
$\ngz$ & $=  \bar T^{3}\, 2\zeta(3)/\pi^{2} $ & background photon number density & \eqref{ex2t} \\ \hline
$u$ &  &fluid velocity potential & \\ \hline
$u^{0}$ & $=\sqrt{1+u_{i}u^{i}}$&  time component of the fluid velocity & \\ \hline
$u_{i}$ & $=\partial_{i} u$ & spacial component of the fluid velocity & \\ \hline
$T^{\mu\nu}$ & & total energy-momentum tensor & \eqref{emt}\\ \hline
$\Delta T^{\mu\nu}$& & viscous corrections to $T^{\mu\nu}$ & \eqref{D} \\ \hline
$N^{\mu}_{x}$ & &number density current of specie ``$x$''  & \eqref{Nperf}\\ \hline
$\Delta N^{\mu}_{x}$ & &viscous corrections to $N^{\mu}_{x}$ & \eqref{D} \\ \hline
$\zeta$ & & bulk viscosity & \eqref{deltaN} \\ \hline
$\chi_{x}$ & & heat conduction of specie ``$x$'' & \eqref{deltaN}\\ \hline
$\eta $ &  &shear viscosity & \eqref{deltaN}\\ \hline
$\delta \rho_{x}$ & $=\rho_{x}-\bar \rho_{x} $ & dimensionful density perturbation & \\ \hline
$\delta_{x}$ & $=\delta \rho_{x}/\bar \rho $ & linear-order dimensionless density perturbation  &  \\ \hline
$\dto$ & $=\delta T_{(1)r}/\bar T$ & linear-order rest-frame temperature perturbation & \eqref{ng}\\ \hline
$\dtt$ & $=\delta T_{(2)r}/\bar T$ & second-order rest-frame temperature perturbation & \eqref{ng}\\ \hline
$\tcu$ & $=\delta T_{(1)}/\bar T$ & linear-order cosmological-frame temperature perturbation & \eqref{ex1} \\ \hline
$\tct$ & $=\delta T_{(2)}/\bar T$ & second-order cosmological-frame temperature perturbation & \eqref{ex1} \\ \hline
$\mu$ & $=-\mth/\bar T$ & $\mu$-distortion parameter & \eqref{mudef}\\ \hline
$\mth$ &  & thermodynamic (relativistic) chemical potential & \eqref{mth} \\ \hline
$\ex{}$ & & quantum/stochastic expectation value & \\ \hline
$\ex{}_{p}$ & $=\int_{t}^{t+2\pi/\omega}$ dt' & time average over one oscillation & \eqref{exbbp}\\ \hline
$\psi$ & $=\frac{1}{2} \left(3 \dot A+\nsq\dot B\right)$ & gravitational perturbation in synchronous gauge & \eqref{psidef} \\ \hline
\end{tabular}
\end{table}
\newpage

\begin{table}[!h]
\centering
\begin{tabular}{|c|l|l|c|}  \hline
Symbol    & Relation &	Meaning 	&  Equation\\ \hline\hline
$\R$ & $= A/2+H u $ & curvature perturbations on comoving hypersurfaces & \eqref{def} \\ \hline
$t_{\gamma}$ & $= \left(n_{e} \sigma_{T}\right)^{-1}$ & photon mean-free path & \eqref{vis}\\ \hline
$\sigma_{T}$ & & Thomson cross section & \eqref{vis} \\ \hline
$q_{D}$ & & diffusion damping scale of $T$ perturbations & \eqref{qD}\\ \hline
$\qmd$ & & diffusion damping scale of $\mu$ perturbations & \eqref{qDm}\\ \hline
$\Gamma_{\chi,\eta}$ & & diffusion damping rate & \eqref{Gamma} \\ \hline
$\ar$ & $=\pi^{2}/15$ & radiation energy constant & \eqref{ar} \\ \hline
$\br$ & $=2\zeta(3)/\pi^{2}$ & radiation number constant & \eqref{ex2t}\\ \hline
$\Ao$ & $=90\zeta(3) / \pi^{4}$ & $\mu$ correction to $\rho_{\gamma}$ & \eqref{ex3} \\ \hline
$\At$ & $=\pi^{2}/[6\zeta(3)]$ & $\mu$ correction to $n_{\gamma}$ & \eqref{ex2t} \\ \hline
$\Atr$ & $=135\zeta(3)/(2\pi^{4})$ & $\mu$ correction to $s_{\gamma}$ & \eqref{As}\\ \hline
\end{tabular}
\caption{Symbols used in the paper. \label{tab}}
\end{table}

 
\section{Introduction}\label{intro}

The cosmic microwave background (CMB) and large-scale structures have allowed us to probe primordial perturbations with very high accuracy in the interval of scales $10^{-4}< k {\rm Mpc}<1$. Very little is known so far for scales that are parametrically smaller $k\gg {\rm Mpc}^{-1}$. The reason is that this information has been a erased by diffusion damping in the CMB temperature anisotropy and is swamped by nonlinear gravitational effects in large-scale structures. 
Fortunately, the frequency dependence of the CMB spectrum does carry primordial information about very small scales. 
Around the time in which the COBE data became available, a considerable effort (see e.g.~\cite{ref1,ref2,ref3,oldbucket}) was devote to the study of the evolution of the CMB spectrum and to the search for potential sources of distortion. This effort has decreased considerably since the COBE collaboration published data compatible with a blackbody spectrum \cite{Fixsen:1996nj}, imposing strong bounds on the distortion parameters, $|y|<1.5 \times 10^{-5}$ and $|\mu|<9 \times 10^{-5}$. No other experiment has since been able to independently improve on COBE/FIRAS constraints. Nevertheless the experiment TRIS \cite{TRIS}, combining its data with the one of COBE/FIRAS was able to slightly tighten the bound on $\mu$, reporting $|\mu|<6 \times 10^{-5}$. Another notable experimental effort was ARCADE \cite{ARCADE}, which detected an anomalous raise in the low-frequency spectrum.
The subject has enjoyed a recent revival \cite{largeb,BE,2t2,Ganc:2012ae,PZ} also thanks to the prospects of an observational detection with future experiments such as PIXIE \cite{PIXIE}. Traditionally, the computations has been performed at the level of the kinetic theory, using Boltzmann transport equation. Given the importance of spectral distortion as probe of primordial perturbations, it is useful to have different but equivalent physical descriptions of the relevant processes. The fluid limit of Boltzmann transport equation is a very intuitive one and has proven to be a good aid to physical intuition in the study of the CMB (for some analytical work emphasizing the fluid dynamic approach see \cite{Weinberg} and references therein). The goal of this paper is to provide a pedagogical, thorough and systematic description of the creation and evolution of $\mu$-type spectral distortion using fluid dynamics. 

A picture of the phenomena that we intend to describe is the following. At early times, $z\ll z_{i}\equiv 2 \times 10^{6}$, thermodynamic equilibrium guarantees that photons are very well described by a blackbody spectrum. Between $z_{i}$ and $z_{f}\equiv 5 \times 10^{4}$, photon number changing processes are frozen out due to the cosmic expansion, but equilibrium is still ensured by elastic Compton scattering. During this era, the dissipation of acoustic modes due to viscosity leads to an increase of the entropy and the energy of the photon-baryon-electron plasma. Since this increase is not accompanied by an appropriate increase in the number of photons, both the average entropy and energy per photon grow. In thermodynamic equilibrium, these are given by
\be
\frac{\rho_{\gamma}}{n_{\gamma}}&=&T 
\frac{\ar}{\br} \left[1+\mu \,
\left(\At-\Ao\right)+ \mathcal{O} \left(\mu^{2}\right)\right] \nonumber\\
&\simeq &2.7\, T \left[1+0.26\,\mu+ \mathcal{O} \left(\mu^{2}\right)\right]\,,\nonumber \\
\frac{s_{\gamma}}{n_{\gamma}}&=& 
\frac{4}{3} \frac{\ar}{\br} \left[1+\mu \,
\left(\At-\Atr\right)+ \mathcal{O} \left(\mu^{2}\right)\right] \label{000}\\
&\simeq &3.6\,  \left[1+0.53\,\mu+ \mathcal{O} \left(\mu^{2}\right)\right]\,,\nonumber
\ee
where $\ar,\,\br,\,\Ao,\,\At,\,\Atr$ are all numerical constants of order one define in table \ref{tab}. Therefore an increase in the entropy per photon is tantamount to the creation of a photon chemical potential\footnote{We use the symbol $\mu$ for the dimensionless size of spectral distortion of the $\mu$ type. This is related to the standard (dimensionful) chemical potential $\mth$ used in thermodynamics by $\mu=-\mth/T$.} $\mu$. In other words, $\mu$ is a direct measurement of how much the rest-frame entropy per photon deviates from the fixed blackbody value. An analogous statement is valid also for the energy per photon but provided one specifies that $T$ is held fixed or some equivalent condition. It is for this reason that $s_{\gamma}/n_{\gamma}$ is a more useful indicator of $\mu$-distortion than $\rho_{\gamma}/n_{\gamma}$.

After $z_{f}$, thermodynamic equilibrium fails to be reached because the plasma temperature has dropped well below the electron mass. From this moment onward, the creation of $\mu$-type distortion is halted (while $y$-distortion can still be generated \cite{ref3}), but the distortion present at the end of the $\mu$-era, $z_{f}$, does survive and evolve in a non-trivial way all the way to us. On large scales, this evolution consists simply of the projection of the inhomogeneities present at $z_{f}$ onto the multipoles of the late time CMB sky. On smaller scales, $l\gtrsim 1000$, diffusion damping and the thickness of the last scattering surface erase any inhomogeneity. This last result is relevant even if we were not to observe these small scales. To understand the reason let us assume a Gaussian universe with just adiabatic initial conditions. Then fluctuations of the $\mu$ generated by the dissipation of acoustic modes ($\ex{\mu}\sim {\rm few }\,10^{-8} $) on large scales are just white noise. Hence the $\mu$-power spectrum on large scales, $C_{l\ll 1000}^{\mu\mu}$, is \textit{linearly} proportional to the largest dissipation scale because of an average along the line of sight. This was derived in \cite{PZ}, where the largest dissipation scale was called $k_{s}$, and will be discussed in section \ref{s:A}. The evolution after the $\mu$-era erases inhomogeneities in $\mu$ also on scales that are larger than the dissipation scale at the end of the $\mu$-era due to the finite thickness of the last scattering surface and further diffusion processes. This suppresses the large scale-power spectrum of $\mu$ even more. Hence, in a Gaussian and ``adiabatic'' universe, there is practically no hope to extract any information from the $\mu$ multipoles. On the other hand, in the presence of primordial non-Gaussianity\footnote{Also in the presence of isocurvature modes things can be more interesting, but we do not discuss this case in the following} things are different \cite{PZ}. $\mu$ is modulated by large-scale potential fluctuations which also produce the large scale temperature anisotropies. Thus its  correlation with $\delta T$ gives a direct measurement of primordial non-Gaussianity. From this point of view, the Gaussian power spectrum discussed above is just some source of noise, typically referred to as cosmic variance. The fact that this cosmic variance is negligibly tiny for $\mu$ anisotropies (generated by the dissipation of acoustic modes) can be of great advantage in constraining or detecting deviations from Gaussianity at very small scales.

 
\subsection{Summary}\label{ss:sum}

Let us give a quick summary of our main results and of the organization of the paper. In section \ref{s:B} we set the stage with a pedagogical review of the hydrodynamic description of the universe much before the $\mu$-era. All the results derived there are already present in the literature. In section \ref{s:D} we adopt the same tools to provide a detailed descriptions of the $\mu$-era and the evolution of the $\mu$-type spectral distortion. A lightning derivation of the main result goes as follows\footnote{This derivation uses rest-frame quantities such as $n_{\gamma}$ and $\rho_{\gamma}$. A parallel derivation using cosmological-frame quantities such as $N^{0}_{\gamma}$ and $T^{00}$ is given in appendix \ref{a:cf}. The two results are identical for $\mu$, since this quantity does not change as we move to a different reference frame. The results for temperature fluctuations, on the other hand, are different but related by the transformation in \eqref{Ttrans}.}. We expand in perturbations around an FLRW universe containing just a photon-electron-baryon plasma\footnote{Dark matter is indeed a small corrections since the $\mu$-era happens entirely during radiation domination. The presence of neutrinos, on the other hand, leads to a $10\%$ error that we fix in the more rigorous treatment in the rest of the paper.}. The evolution equations are the conservation of the total energy momentum tensor $T^{\mu\nu}_{;\nu}=0$, the conservation of baryon number $N^{\mu}_{b;\mu}=0$ and, during the $\mu$-era, the conservation of photon number $N^{\mu}_{\gamma ;\mu}=0$. At linear order in perturbations these equations describe the propagation of waves. Including dissipative corrections, the waves get damped once they reach some small scale $q_{D}$. Perturbations to the photon density $\delta_{\gamma}$ and velocity potential $u$ (so that the irrotational fluid velocity is given by $u_{i}=\partial_{i}u$) are of the form
\be
 \frac{1+w}{c_{s}}\, q \,u(t,q) \simeq \delta_{\gamma}(t,q)\simeq A \cos \left(\int \frac{q}{a}c_{s} dt'\right) e^{- \int \Gamma dt' }\,,\label{stu}
 \ee
where $A$ is some amplitude determined by the superhorizon primordial perturbations, $c_{s}^{2}\simeq w\simeq 1/3$ are the speed of sound of the fluid and the background expansion index, respectively, and $\Gamma$ is the dissipation rate given explicitly in \eqref{Gamma}. There are corrections of order the baryon loading $R$, which we neglect in this summary, but will be accounted for in the following sections. In any case, these corrections are small, since during the $\mu$-era $R\ll 10^{-2}$.

To study the evolution of $\mu$ we first notice that according to \eqref{000}
\be
\partial_{t} \left(\frac{s_{\gamma}}{n_{\gamma}}\right)=\dot \mu 
\frac{4}{3}\frac{\ar}{\br} \left(\At-\Atr \right)\simeq 1.9\, \dot \mu\,.\label{mudsum}
\ee
Since according to COBE/FIRAS \cite{Fixsen:1996nj} $\mu\ll 1$, here and in the rest of the paper we work at linear order in $\mu$. To compute $\dot{s_{\gamma}}$ and $\dot{n_{\gamma}}$ we use the conservation of photon number and the evolution of the entropy density
\be
N^{\mu}_{\gamma;\mu}&\equiv& \left(n_{\gamma}u^{\mu}+\Delta N^{\mu}_{\gamma}\right)_{;\mu}=0\,,\label{nsum}\\
s^{\mu}_{;\mu}&\equiv& \left(s u^{\mu}+\Delta s^{\mu}_{\gamma}\right)_{;\mu}=-\frac{1}{T}\Delta T^{\mu\nu} u_{\mu;\nu}+\Delta N^{\nu}\partial_{\nu}\mu\,,\label{ssum}
\ee
where the latter equation is derived in appendix \ref{a:s}. Here $u^{\mu}$ is the fluid velocity of energy transport (see e.g.~section \ref{s:B}) while $\Delta N^{\mu}$ and $\Delta s^{\mu}$ are viscous corrections that account for the fact that, for an imperfect fluid, the velocities of the transport of photon number and entropy can be different from $u^{\mu}$. As we show in appendix \ref{a:chi}, up to terms of order $R$, $\Delta N^{\mu}_{\gamma}=0$, so we can focus on $\Delta T^{\mu\nu}$. Its general form at leading order in derivatives is given in \eqref{deltaT} and it is proportional to the shear viscosity $\eta$, which for the photon-electron-baryon fluid takes the value \cite{dod} $\eta=(16/45) t_{\gamma}\rgz$, where $t_{\gamma}$ is the photon mean free path. Solving\footnote{This is most easily done in the rest frame of the fluid where $\partial_{\mu}u^{0}=u^{i}=0$ and $u^{0}=1$.} \eqref{nsum} and \eqref{ssum} for $\dot{s_{\gamma}}$ and $\dot{n_{\gamma}}$ one finds
\be
\partial_{t} \left(\frac{s_{\gamma}}{n_{\gamma}}\right)=-\frac{1}{nT} \Delta T^{\mu\nu}u_{\mu;\nu}\,.\label{sumint}
\ee
Since the left hand side is proportional to $\dot \mu$, this implies that $\mu$-distortion is created only when the viscous corrections are important (again this is true only up to corrections of order $n_{b}/n_{\gamma}$). As discusses in subsection \ref{ss:bzo}, $\Delta T^{\mu\nu}$ starts at linear order in perturbations and $\Delta T^{0\mu}_{(1)}=\Delta T^{ii}_{(1)}=0$. Then the leading term in the right hand side of \eqref{sumint} is second order in the small amplitude of perturbations. The explicit computation using \eqref{mudsum}, \eqref{DTij} and \eqref{sumint} gives
\be
\boxed{
\dot \mu=\frac{4}{15 (\At-\Atr)}t_{\gamma} \left[ \left(\partial_{i}u_{j}\right) \left(\partial^{i}u^{j}+\partial^{j}u^{i}\right)-\frac{2}{3} \left(\partial_{l}u^{l}\right)^{2}\right]\,. }
\ee
If we take the ensemble average, focusing on the average $\mu$-distortion, and use the relation between $u$ and $\rgo$ \eqref{stu} and the explicit expression for $\Gamma$ \eqref{Gamma}, this expression agrees with the result of \cite{HSS}
\be
\ex{\dot{ \mu}}&=& -1.4 \partial_{t} \ex{(E_{w}/\rho)}\\
&=&-1.4 \, \frac{c_{s}^{2}}{1+w} \partial_{t} \ex{\ex{\delta_{\gamma}^{2}}}_{p}\,,
\ee
where $E_{w}$ is the energy of a wave, $\ex{}$ the expectation value over the ensemble of all histories and $\ex{}_{p}$ is the average over a period. Notice however that it is important to include the correct relativistic factor $(1+w)^{-1}\simeq 3/4$ (derived in appendix \ref{a:cs}) in the formula for $E_{w}$. Therefore the derivation of the correct numerical factor (first noticed in \cite{2t2}) can be obtained entirely within the realm of fluid dynamics, without any reference to the microscopic theory. 

In the more detailed description of section \ref{s:D} we identify other subleading contributions to $\mu$-distortion:
\begin{itemize}
\item The baryon temperature leads to adiabatic cooling (subsections \ref{s:Bin} and \ref{ss:Dzo}) \cite{BE} and a non-vanishing but small bulk viscosity (subsection \ref{Bvf}).
\item The baryon energy density (mostly rest mass energy) leads to a small damping of the amplitude of acoustic waves (subsections \ref{ss:Bpf}, \ref{ss:Ds} and \ref{ss:Dsv}) and is small during the $\mu$-era, but could be relevant for the $y$-era.
\item Similarly to what happens for baryons, the conservation of photon number is corrected by heat conduction (derived in appendix \ref{a:chi}).
\end{itemize} 

After the end of the $\mu$-era, the dissipation of acoustic modes cease to produce spectral distortion of the $\mu$-type. Nevertheless, in order to make predictions for late-time observables, we need to evolve the signal up to the last scattering surface and then account for the free streaming of photons all the way to us. We do this in section \ref{s:A}. The final result \eqref{ft}, is simple to understand. $\mu$-distortion, defined through the dimensionless chemical potential $\mu$, is not affected by free streaming or gravitational potentials. Hence for $l<1000$ the transfer function is just the simple projection of the $\mu$ monopole at the last scattering surface onto various multipoles as in \eqref{proj}. For $l\gtrsim 1000$ two effects become important: the finite thickness of the last surface \eqref{ft} and diffusion damping \eqref{dd}. The outcome is an exponential damping of $\mu$ anisotropies for $l\gtrsim 1000$, and a suppression of its power spectrum, as discusses in section \ref{s:A}.

\section{Before the $\mu$-era}\label{s:B}

To set the stage, in this section we describe the evolution of the universe at early times, $z\gg z_{i}\equiv2\times 10^{6}$, when double Compton and Bremssstrahlung interactions of photons with the electron-baryon plasma were very efficient. At that time almost perfect thermodynamical equilibrium was maintained and the photon occupation number was very close to a black body spectrum. Since photons can be efficiently produced, the only conserved number in this \textit{black-body era} is the baryon number. After introducing the relevant hydrodynamical equations, we derive the evolution of primordial adiabatic perturbations at linear order and present those second order result that are relevant for the production of spectral distortion.

Because of the frequent interactions, photons electrons and baryons combine into a single fluid, whose energy momentum tensor is covariantly conserved
\be\label{T}
T^{\mu\nu}_{;\mu}=0\,.
\ee
These equation have to be supplemented by the conservation equations of any conserved charge in the system.
After baryogenesis, when the temperature has dropped well below a GeV, the baryon number is conserved. Later, after electron positron annihilation, $z\sim 10^{9}$, also the electron number becomes effectively conserved. However ti a very good approximation matter behaves as a single fluid, so it suffices to follow baryon number conservation
\be\label{N}
N_{b\,;\mu}^{\mu}=0\,,
\ee
where $N^{\mu}$ is the baryon-number current, defined such that in the rest frame of the fluid $N^{0}$ corresponds to the baryon-number density. Finally we need to consider Einstein Equations
\be
G_{\mu\nu}=- \Mpl^{-2} T_{\mu\nu}\,,
\ee
where $\Mpl\equiv \left(8\pi G_{N}\right)^{-1/2}$ is the reduced Planck mass.


\subsection{Inviscid fluid}\label{s:Bin}

The energy-momentum tensor for a perfect fluid is diagonal and isotropic in the rest frame of the fluid, hence in any frame it is given by
\be\label{emt}
T_{\mu\nu}= \left(\rho+p\right)u_{\mu}u_{\nu}+g_{\mu\nu} p\,,
\ee
where $g_{\mu\nu}$ is the metric, $u^{\mu}$ is the four-velocity of a fluid element with $u_{\mu}u^{\mu}=-1$, the energy density is the one measured in the rest frame of the fluid $\rho\equiv u_{\mu}u_{\nu}T^{\mu\nu}$ and for a barotropic fluid, the equation of state gives the pressure $p$ as function of $\rho$. As mentioned previously, in the presence of a conserved quantity, such as baryon number, there is a Lorentz frame in which $N^{\mu}=\{n,\vec 0\}$, with $n$ the number density in the rest frame. For a perfect fluid, the energy-momentum tensor is diagonal in this \textit{same} Lorentz frame, hence in any frame we can write
\be\label{Nperf}
N^{\mu}=n u^{\mu}\,,
\ee
so that $n= - u_{\mu}N^{\mu}$. For baryons the pressure, number density and energy density are related by\footnote{We set the Boltzmann constant to one, $k_{B}=1$, i.e. measure temperature in units of energy.} 
\be\label{bed}
p_{b}=n_{b} T\,,\quad \rho_{b}=\frac{3}{2}m n_{b}+T n_{b}\,.
\ee 
Here $m\simeq$ GeV is the average mass per baryon and $T$ is the baryons temperature, which until late times, $z\sim 200$, differs little from the photon temperature. We will neglect this difference and use $T$ for the whole photon-baryon-electron fluid. For photons, the pressure is given by
\be
p_{\gamma}=\rho_{\gamma}/3\,, 
\ee
independently on whether the photon number is conserved or not. For a black body spectrum
\be\label{SB}
\rho_{\gamma}=\ar T^{4}\,,\label{ar}\,,
\ee
with $\ar\equiv\pi^{2}/15$. For the photon-baryon-electron fluid we then take
\be
\rho=\rho_{\gamma}+m n_{b}+T n_{b}\,,\quad p=\frac{1}{3}\rho_{\gamma}+n_{b}T\,.
\ee
Notice that baryon temperature is typically a small correction. On the one hand, as long as we are interested in the time much after big bang nucleosynthesis (BBN), $T\ll .1$ MeV, and so we can neglect the temperature dependent term in the baryon energy density \eqref{bed}, up to corrections of the order $T/m\ll10^{-4}$. On the other hand
\be\label{r}
  r(t) \equiv\frac{p_{b}}{\rho_{\gamma}}\simeq\frac{\nbz \Tz}{\rgz}=\frac{\nbz}{\ngz} \frac{30 \zeta(3)}{\pi^{4}}\simeq 0.37 \times \frac{\nbz}{\ngz}\simeq 2.2\times 10^{-10}\,,
\ee
with $\zeta$ the Riemann zeta function and $n_{\gamma}$ the number density of photons. Despite being time dependent, the number of photons per baryon, which we parameterize by $r$, has changed very little since electron-positron annihilation. BBN and CMB bounds agree, within the error bars, on the value $\nbz/\ngz\simeq 6\times 10^{-10}$.

Because its effects are suppressed by these small factors, baryon temperature can play an important role only when all other contributions cancel precisely. This does indeed happen for the spectral distortion generated by the homogeneous adiabatic evolution during the $\mu$-era, as we will review in sections \ref{Bv} and \ref{Dv}. In all other sections but these two, we can safely neglect terms proportional to the baryon temperature.

We now proceed to solve \eqref{T} and \eqref{N} in a perturbative expansion around a flat FLRW universe. It will be convenient to introduce the notation\footnote{\label{foot}Some authors (e.g.~\cite{Weinberg}) define $\tilde \delta_{\gamma}\equiv \delta \rho_{\gamma}/(\bar \rho_{\gamma}+\bar p_{\gamma})$. Our choice of $\delta$ is related to that choice by
\be
\rgo=\frac{(1+R)4}{3+4R}\tilde \delta_{\gamma}\label{conv}
\ee
where $R\equiv 3\rbz/(4\rgz)$ is the baryon loading.}
\be
\rho_{b,\gamma}&=&\bar \rho_{b,\gamma}\left[1+\ru_{b,\gamma}+\rt_{b,\gamma}\right] \,,\label{dim}\\
 u_{i}&=&\bar u_{i}+\uu+\ut\,,\\
 n_{b,\gamma}&=& \bar n_{b,\gamma} \left[1+\no_{b,\gamma}+\nt_{b,\gamma}\right]\label{less}
\ee
so that any $\delta$ and $\no$ are dimensionless. A bar indicates zeroth order homogeneous (but time dependent) quantities and to simplify the notation we omit label ``$(1)$'' in the first order perturbations. Finally notice that due to the tight coupling between photons and electron-baryons, there is a single fluid velocity $u_{i}$. Since vector perturbations decay in and FLRW, we neglect vorticity and introduce the velocity potential $u$ by
\be
u_{i}=\partial_{i}u\,.
\ee


\subsubsection{Zeroth order}\label{Bv}

We consider a homogeneous and isotropic background, so that $\rho=\bar \rho(t)$ and $\bar u_{i}=0$, and the unperturbed metric is the one of an FLRW universe
\be
ds^{2}=-dt^{2}+a^{2}d\vec x^{2}\,,\label{FLRW}
\ee 
with $a(t)$ the scale factor. The Christoffel symbols are found to be
\be\label{cs}
\Gamma^{i}_{j0}=\Gamma^{i}_{0j}=H\delta_{ij}\,,\quad \Gamma^{0}_{ij}=a^{2}H\delta_{ij}\,,
\ee
where $H(t)\equiv\dot a/a$ is the time-dependent Hubble parameter. Einstein equations then reduce to the Friedamann equation
\be
3\Mpl^{2}H^{2}=\rgz+\rbz+\bar \rho_{c}+\bar\rho_{\nu}\,,
\ee
where $\bar\rho_{c,\nu}$ are the dark matter and neutrino contributions, respectively. Using \eqref{cs}, the number \eqref{N} and energy conservations \eqref{T} lead to
\be
\nbzd&=&- 3 H \nbz\,,\\
\rgzd&=&-4 H \rgz \left[1+\frac{3}{8} r+\mathcal{O} \left(r^{2}\right)\right]\,,\label{rgz}
\ee
where $r$ was defined in \eqref{r}. Since there are about $10^{9}$ photons per baryon, the baryon pressure can be neglected up to corrections of order $r\simeq 2.2\times 10^{-10}$. In \eqref{rgz} we see that, the effect of the baryon temperature is to accelerate the dilution of the radiation energy density. This can be understood (see e.g.~\cite{Weinberg}) by recalling that if the photons and the baryon were not interacting, their temperature would redshift as $a^{-1}$ and $a^{-2}$, respectively. Once the interaction is taken into account, the common temperature scales with a power of $a$ in between $-1$ and $-2$. Since the are so many more photons than baryons, the scaling in actually very close to $-1$, as can be seen by rewriting \eqref{rgz} using \eqref{SB}
\be\label{Tbm}
\Tzd&=&- H \Tz \left[1+\frac{3}{8} r+\mathcal{O} \left(r^{2}\right)\right]\,.
\ee
%


\subsubsection{First order}\label{ss:Bpf}

For the study of perturbations we choose to work in synchronous gauge\footnote{We use the notation of \cite{Weinberg}, except for the definition of the density perturbation as explained in footnote \ref{foot}}
\be\label{mper}
ds^{2}=-dt^{2}+a^{2}dx^{i}dx^{j}\left[ \left(1+A\right)\delta_{ij}+B_{,ij}\right]\,,
\ee
and define
\be\label{psidef}
\psi\equiv\frac{1}{2} \left(3 \dot A+\nsq\dot B\right)\,.
\ee
With this choice, \textit{at linear order} the time component of the velocity vanishes $\delta u^{0}_{}=\delta g_{00}/2=0$ and no metric perturbation appears in $u_{i}=a^{2}u^{i}$. Also the only non-vanishing metric perturbations are $\delta g^{ij}=- a^{-4} \delta g_{ij}$, so indices of metric perturbations are \textit{not} raised and lowered by the background metric.

Let us now expand equations \eqref{T} and \eqref{N} at linear order in perturbations
\be
\partial_{t} T^{00}_{(1)}+3HT^{00}_{(1)}+a^{2}HT^{ii}_{(1)}+\partial_{i}T^{i0}_{(1)}&=&- T^{0\alpha}_{(0)} \delta\Gamma^{\mu}_{\mu\alpha}-T^{\mu\alpha}_{(0)}\delta\Gamma^{0}_{\mu\alpha}\,,\label{00}\\
\partial_{t} T^{i0}_{(1)}+5HT^{i0}_{(1)}+\partial_{j}T^{ij}_{(1)}&=&- T^{i\alpha}_{(0)} \delta\Gamma^{\mu}_{\mu\alpha}-T^{\mu\alpha}_{(0)}\delta\Gamma^{i}_{\mu\alpha}\,,\label{i}\\
\partial_{t} N^{0}_{(1)}+3HN^{0}_{(1)}+\partial_{i} N^{i}_{(1)}+\psi N^{0}_{(0)}&=&0\label{n}\,.
\ee
Neglecting the baryon temperature, i.e. terms of order $r$, the energy-momentum tensor at linear order is given by
\be
T^{00}_{(1)}&=&\rgz  \left(\rgo+\frac{4}{3}R\rbo\right)\,,\\
T^{0i}_{(1)}&=&\frac{4}{3}\rgz (1+R) u^{i}\,,\\
T^{ij}_{(1)}&=&\frac{1}{3} \rgz a^{-2} \left(\delta_{ij} \rgo-a^{-2} A \delta_{ij}-a^{-2}B_{,ij}\right)\,,
\ee
and $\delta \Gamma$ can be found in textbooks, e.g.~\cite{Weinberg}. Going to Fourier space, (\ref{00}-\ref{n}) become
\be
\frac{3}{4} \rgod+ R \rbod- \frac{q^{2}}{a^{2}} (R+1) u&=& - (R+1) \psi \,,\label{01}\\
(R+1) \dot u- H u + \frac{1}{4}\rgo&=&0\,,\label{i1}\\
\nbod-\frac{q^{2}}{a^{2}}u&=&-\psi\,,\label{n1}
\ee
where $R\equiv 3\rbz/(4\rgz)$ is the baryon loading, which to leading order is just proportional to the scale factor $a$. While Einstein equations give \cite{Weinberg}
\be\label{psi}
\partial_{t} \left(a^{2}\psi\right)=-\frac{1}{2} \left(2\rgz \rgo+ \rbz \rbo +\bar \rho_{c} \delta_{c}+2 \bar\rho_{\nu} \delta_{\nu}\right)\,.
\ee
One can solve \eqref{n1} for $\nbod=\rbod$, \eqref{01} for $u$, \eqref{i1} for $\dot u$ and plug them into the time derivative of \eqref{01}. The result is
\be\label{ddd}
\rgodd+H\frac{1+2R}{1+R} \left[\rgod+\frac{4}{3} \psi\right]+\frac{q^{2}}{a^{2}} c_{s}^{2} \rgo+\frac{4}{3}\dot\psi=0\,,
\ee
where we have introduced the speed of sound in the photon-baryon-electron fluid $c_{s}^{2}\equiv \left[3 \left(1+R\right)\right]^{-1}$. Analytical solutions of \eqref{psi} and \eqref{ddd} can be quite involved, but as long as we are interested in the evolution of perturbations well inside the horizon things are much simpler. During radiation domination, the gravitational perturbation $\psi$ decays inside of the horizon as $q^{2}/(aH)^{2}$ due to the large pressure of photons, so we can drop all $\psi$ terms in \eqref{ddd}. Then \eqref{ddd} can be easily solved by the WKB approximation. 

We still need to find the correct normalization of $\rgo$ fluctuations inside the horizon in terms of primordial superhorizon fluctuations. Here we consider only adiabatic initial conditions, specified in terms of the gauge invariant variable $\R$ defined in \eqref{def}, which in the simplest cosmological scenarios is conserved outside the horizon. If we neglect dark matter and neutrino, then the universe is filled with a single fluid and can be described using a $P(X)$ Lagrangian as in \cite{}. The derivation of the initial condition, whose details are collected in appendix \ref{a:R}, then proceeds as follows. One works with the gauge invariant variable $\R$ whose action \eqref{Ract} is well know. Its equations of motion can be easily solved both inside and outside the horizon, during radiation domination. Then well inside the horizon one goes to synchronous gauge in which $\R$ and the velocity potential $u$ are directly related. Finally one can relate this solution for $u$ to a solution for $\rgo$ using \eqref{01}. The final result is
\be
\rgo(t)&=&- \frac{4\R^{0}}{(1+R)^{1/4}}\,\cos \left(\int^{t} \frac{q}{a(t')}c_{s}(t')\,dt'\right)+\mathcal{O} \left(\frac{\bar\rho_{\nu}}{\rgz}\right)\,,\label{prima}
\ee
with $\R^{0}$ the asymptotic value of $\R$ at early times, outside of the horizon, which is determined by the microphysics of the early universe, e.g.~by inflation. The factor $(1+R)^{-1/4}$ is best understood\footnote{One might be concerned that terms of order $R$ were dropped in computing the normalization of $\rgo$ in terms of the conserved quantity $\R$. Notice however that $R$ grows with time, so it makes sense to neglect terms of order $R_{\rm early}$ and keep terms of order $R_{\rm late}\gg R_{\rm early}$.} by computing the leading order WKB solution of \eqref{ddd}.

The above derivation does not capture the dark matter and neutrino contributions. While the first component is indeed very small during radiation domination, the second leads to a $10\%$ correction. The numerical analysis of \cite{Bashinsky} gives
\be
\rgo(t)&=&-\frac{4\R^{0}}{(1+R)^{1/4}}\, \left(1-0.268 R_{\nu}\right)\,\cos \left(\int^{t} \frac{q}{a(t')}c_{s}(t')\,dt'\right)\,.
\ee
Here and in \eqref{prima} we have re-absorbed a neutrino dependent phase into the lower bound of the integral. The quantities we are interested in contain the average of oscillations over a period, hence the phase will be irrelevant for us. Here we have defined ratio of neutrino to photon energy density
\be\label{Rnu}
R_{\nu}\equiv \frac{\bar\rho_{\nu}}{\rgz}=N_{\nu} \frac{7}{8} \left(\frac{4}{11}\right)^{4/3}\,.
\ee
For the standard model with three neutrinos, $N_{\nu}=3.04$ and one finds $R_{\nu}=0.41$. Combining \eqref{01} and \eqref{n1}, one finds
\be\label{usq}
\frac{q^{2}}{a^{2}}u=\frac{3}{4}\rgod+\psi\,.
\ee
If we focus on modes well inside the horizon and neglect $\psi$, to leading order in $aH/q$ we get
\be
\ex{u^{i}u_{i}}= \int \frac{d^{3}q}{(2\pi)^{3}}\frac{ |\R^{0}(q)|^{2}}{(1+R)^{1/2}}\left[3 c_{s}   \left(1-0.268 R_{\nu}\right)\,\sin \left(\int^{t} \frac{q}{a}c_{s}\,dt'\right) \right]^{2}\,,\label{wont}
\ee
where by statistical isotropy and homogeneity
\be
\ex{\R^{0}(\vec q) \R^{0}(\vec q')^{\ast}}=(2\pi)^{3} \delta^{3} \left(\vec q+\vec q'\right) |\R^{0}(q)|^{2}\,.
\ee
For future reference we also compute
\be
\ex{\rgo^{2}}= \int \frac{d^{3}q}{(2\pi)^{3}}\frac{ |\R^{0}(q)|^{2}}{(1+R)^{1/2}}\left[4   \left(1-0.268 R_{\nu}\right)\,\cos \left(\int^{t} \frac{q}{a}c_{s}\,dt'\right) \right]^{2}\,,\label{dont}
\ee
A few comments are in order. First, taken at face value, given a scale invariant spectrum of primordial perturbations both integrals in \eqref{wont} and \eqref{dont} are logarithmically divergent (in the UV and IR). We will see that for the physically relevant quantity, i.e.~$\mu$, the divergences are naturally cut off. Second, notice that the velocity potential and density perturbations oscillate with a phase difference of $\pi/2$. Hence, if we denote by $\ex{}_{p}$ the time average over one oscillation (while $\ex{}$ is still the quantum expectation value) we find
\be
\ex{\ex{u^{i}u_{i}}}_{p}= \left(\frac{3 c_{s}}{4}\right)^{2}\ex{\ex{\rgo^{2}}}_{p}= \left[\left(\frac{3 c_{s}}{4}\right)^{2}\ex{\rgo^{2}}+\ex{u^{i}u_{i}}\right]\frac{1}{2}\label{exbbp}
\ee
Finally, the amplitude of density and velocity waves, i.e.~$\ex{\ex{u^{i}u_{i}}}_{p}$ and $\ex{\ex{\rgo^{2}}}_{p}$, depends very weakly on time. Deviations from a constant amplitude come at order $R$, which is very small during the $\mu$-era. We will come back to this point at the end of subsection \ref{ss:Ds}.


\subsubsection{Second order}\label{ss:sB}

Let us move on to second order perturbation theory. A thorough study of second-order perturbation theory is beyond the scope of this work. We will instead concentrate on the results relevant for the production of spectral distortion, i.e.~we want to study how the dynamics of short waves feeds back at quadratic order into the homogeneous evolution. First of all we will neglect gravity perturbations. The reason is that we will be interested in the physics of perturbations that are well inside the horizon where gravitational potentials are negligible. A great simplification is achieved by studying the evolution of the expectation values $\ex{}$ of the various perturbations, rather than the perturbations themselves. Assuming statistical homogeneity and isotropy one can then discard all derivative terms.

Let us take the expectation value\footnote{This can be thought of as a quantum/statistical expectation value or as an average over space.} of the conservation of energy $\ex{T^{0\mu}_{(2);\mu}}=0$ and particle number $\ex{N^{\mu}_{b;\mu}}=0$. Spatial derivatives can be taken out of the expectation value. Then invoking statistical isotropy they must all vanish. One is left with:
\be\label{fs}
\ex{ \left(3H+\partial_{t}\right)T^{00}_{(2)}+a^{2}H  T^{ii}_{(2)} }&=&0\,,\\
\ex{\left(3H+\partial_{t}\right)N^{0}_{(2)}}&=&0\,,
\ee
where
\be
T^{00}_{(2)}&=& \frac{4}{3}\rgz \left[\frac{3}{4}\rgt+(1+R) u_{i}u^{i}+R \rbt\right]\,,\\
T^{ij}_{(2)}&=&\frac{4}{3}\rgz \left[a^{-2}\delta_{ij} \frac{1}{4}\rgt+ (1+R) u^{i}u^{j} \right]\,,\\
N^{0}_{(2)}&=& \nbz \nbt + \frac{1}{2} u^{i}u_{i}\,,
\ee
and we used $u^{0}_{(2)}=u^{i}u_{i}/2$. We focus on the ensemble average and use statistical homogeneity drop all total space derivatives. Then
\be
\ex{\frac{3}{4}\rgtd+(1+R)  \partial_{t}\left(u_{i}u^{i}\right)+R \rbtd+RHu_{i}u^{i}}&=&0 \,,\nonumber\\
\partial_{t} \left(\ex{\rbt+\frac{1}{2} u_{i}u^{i}}\right)&=&0\,.\label{sBN}
\ee
We differ a discussion of this result to subsection \ref{ss:Ds}, where we will be able to include the effects of diffusion damping.


\subsection{Viscous fluid}\label{Bvf}

The perfect fluid provides just the zeroth order behavior in an expansion in small momenta. Every realistic fluid deviates from this idealized case and deviations can be systematically parameterized in terms of viscous coefficients. For our purposes it will suffice to consider just the first order corrections in the derivative expansion. In general one will have (see e.g.~\cite{Weinberg0,Weinberg})
\be\label{D}
T^{\mu\nu}\rightarrow T^{\mu\nu}+\Delta T^{\mu\nu}\,,\quad N^{\mu}\rightarrow  N^{\mu}+\Delta N^{\mu}\,.
\ee
where $T^{\mu\nu}$ and $N^{\mu}$ are still given by the perfect fluid formulae \eqref{emt} and \eqref{Nperf}. One can still define $\rho$ and $n$ as the energy and number density in the rest frame defined by $u^{\mu}$, i.e.~$\rho\equiv u_{\mu}u_{\nu} \left(T^{\mu\nu}+\Delta T^{\mu\nu}\right)$ and $n\equiv - u_{\mu} \left(N^{\mu}+\Delta N^{\mu}\right)$. This implies that in any frame 
\be
u_{\mu}u_{\nu} \Delta T^{\mu\nu}=0\,,\quad  u_{\mu} \Delta N^{\mu}=0\,.
\ee
Then one can still redefine $u^{\mu}$ such that \textit{either} $T^{0i}+\Delta T^{0i}$ \textit{or} $N^{i}+\Delta N^{i}$ vanish in the rest frame defined by $u^{\mu}$. We choose the first option so that in any frame $u_{\mu}\Delta T^{0\mu}=0$. In words, this means that $u^{\mu}$ is the velocity of energy transport. Because of this choice the coarse grained fluid velocity of particles whose number is conserved, e.g.~baryons, is not the velocity $u^{\mu}$. Hence their conservation has to be corrected by a non-vanishing $\Delta N^{\mu}$. The generic corrections can be parameterized by three coefficients according to 
\be
\Delta T_{\mu\nu}&=&-\eta \left[u_{\nu;\mu}+u_{\mu;\nu}+u_{\nu}u^{\gamma}u_{\mu;\gamma}+u_{\mu}u^{\gamma}u_{\nu;\gamma}\right]\nonumber \\
&&\quad + \left(\frac{2}{3}\eta -\zeta\right) u^{\gamma}_{;\gamma}\left(g_{\mu\nu}+u_{\mu}u_{\nu}\right)\,,\label{deltaT} \\
\Delta N^{\mu}&=& -\chi \left(\frac{nT}{\rho+p}\right)^{2} \left[\partial_{\mu} \mu+u^{\mu}u^{\gamma}\partial_{\gamma} \mu\right]\,,\label{deltaN}
\ee
where $\eta$, $\zeta$ and $\chi$ are the shear viscosity, bulk viscosity and heat conduction respectively. The second law of thermodynamics requires them to be all non-negative. In \eqref{deltaN} we have defined the dimensionless chemical potential $\mu$ by dividing the standard chemical potential by $T$. This definition agrees with the common use in the literature on spectral distortion and is quite convenient for photons, as we will see.

The viscous coefficients $\eta$, $\zeta$ and $\chi_{b}$, where $\chi_{b}$ is the heat conductivity appearing in the conservation of baryon number have been computed\footnote{The shear viscosity in \cite{Weinbergold} should be corrected for the contribution of photons polarization \cite{Weinberg}.} in \cite{Weinbergold}
\be
\eta=\frac{16}{45} \rgz t_{\gamma}\,, \quad \chi_{b}T=\frac{4}{3}\rgz t_{\gamma}\,,\quad \zeta=4 \rgz t_{\gamma} \left[\frac{1}{3}- \left(\frac{\partial p}{\partial \rho}\right)_{n}\right]^{2}\,.\label{vis}
\ee
Here $t_{\gamma}\equiv 1/(\sigma_{T}n_{e})$ is the photon mean free path, with $n_{e}$ the density of free electrons of mass $m_{e}$ and charge $e$ and $\sigma_{T}=8\pi e^{4}/ (3 m_{e}^{2})\simeq 66\, {\rm fm}^{2}$ is the Thomson cross section. In appendix \ref{a:chi} we rederive $\chi_{b}$ as well as the analogous heat conduction for photons $\chi_{\gamma}$ which is relevant during the $\mu$-era, when the number of (not-so-soft) photons is conserved. The bulk viscosity is extremely small but non-vanishing because of the baryon temperature. From \eqref{vis} one finds
\be\label{zeta}
\zeta= \rgz\, t_{\gamma}\, \frac{r^{2}}{9}+\mathcal{O}\left(r^{3}\right)\ll T \chi_{b},\,\eta\,.
\ee


\subsubsection{Zeroth Order}\label{ss:bzo}

As in the perfect fluid case, we choose a homogeneous and isotropic background with $\rho=\rz$ and $u_{i}=0$. By the definition of $\rho$ and $u^{\mu}$ one must have $\Delta T^{0\mu}_{(0)}=0=\Delta N^{0}_{(0)}$. By isotropy $\Delta N^{i}_{(0)}=0$ and the only non vanishing component is then the spatial diagonal part
\be
\Delta T^{ij}=-\zeta a^{-2} \delta_{ij} 3H\,.
\ee
We see that even at the zeroth order, entropy could be created \cite{Weinbergold} in the presence of a sizable bulk viscosity. Since $\zeta$ for the photon-baryon-electron plasma is so small (see \eqref{zeta}), we can safely neglect this effect, up corrections of order $r^{2}\sim 10^{-19}$. In conclusion the background evolution is unaltered from our perfect fluid discussion of subsection \ref{Bv}. 

Before proceeding, let us see which other constraints can be derived on the viscous contribution. By using 
\be
u^{0}_{(0)}=1\,, \quad u^{0}_{(1)}=0\,, \quad u^{i}_{(0)}=0\,,\quad \Delta T^{\mu\nu}_{(0)}=0\,,\quad \Delta N^{\mu}_{(0)}=0\,,
\ee
and $u_{\mu}\Delta T^{\mu\nu}=0=u_{\mu}\Delta N^{\mu}$ at subsequent orders in perturbation theory, one finds
\be\label{vic}
\Delta T^{0\nu}_{(1)}&=&0\,,\quad \Delta T^{00}_{(2)}=0\,,\quad \Delta T^{i0}_{(2)}=u_{j,(1)}\Delta T^{ij}_{(1)}\,, \nonumber\\
\Delta N^{0}_{(1)}&=&0\,,\quad \Delta N^{0}_{(2)}=u_{i,(1)}\Delta N^{i}_{(1)}\,.\label{did}
\ee


\subsubsection{First order}\label{ss:Bfv}

At linear order in perturbations, we want to compute $\Delta T^{\mu\nu}_{;\nu}$. These viscous terms in general become relevant only at some dissipation scale which is approximately given by the geometric mean between the Hubble and mean free path scales. Physically this corresponds to distance that a typical microscopic particle covers due to Brownian motion in a time $H^{-1}$ \cite{dod}. Given that $t_{\gamma}\ll H$ at any time before decoupling, the dissipation scale is well inside the horizon. During radiation domination the gravitational potentials decay inside the horizon as $q^{2}/(aH)^{2}$, so we safely neglect them in computing $\Delta T$ and $\Delta N$. In order to use \eqref{deltaN} we need to compute the chemical potential for baryons. This is straightforward once we have any one thermodynamical potential in terms of its natural variables. A simple choice is to compute the entropy from the microcanonical potential (Sackur-Tetrode equation). Ignoring very small quantum effects one finds\footnote{Notice that the relativistic chemical potential is related to the non-relativistic one by $\mth=\mu_{\rm non-rel}+m$. This is because in the non-relativistic description the energy does not contain the rest mass of the particle.}
\be\label{mub}
\frac{\mthb}{T}= \left[\frac{\partial S(U,V,N)}{\partial N}\right]_{U,V}=\frac{m}{T}+  \log \left[\frac{n}{ (m T)^{3/2}}\right]+\frac{3}{2}\log \left(2\pi \right) \,.
\ee
Given that after BBN $T\ll m$, one can neglect the logarithmic contributions. Using \ref{vic} and neglecting baryon temperature, i.e.~corrections of order $r$, we find
\be
\Delta T^{0\nu}_{;\nu}&=&a^{2}H\Delta T^{ii}_{(1)}=0\,,\label{eq0}\\
\Delta T^{i\nu}_{;\nu}&=&\partial_{j}\Delta T^{ij}_{(1)}\\
&=&\partial_{j} \left[-\eta a^{-2}\left(\partial_{j}u^{i}+\partial_{i}u^{j}-\frac{2}{3}\delta_{ij}\partial_{l} u^{l}\right)-\zeta a^{-2}\delta_{ij}\partial_{l} u^{l}\right]\,,\label{DTij}\\
\Delta N^{\mu}_{;\mu}&=& a^{-3}\partial_{\mu} \left(a^{3} \Delta N^{\mu}\right)\\
&=&\nbz\frac{\chi_{b}\bar T}{\frac{4}{3}\rgz} \frac{R}{(1+R)^{2}} \left[\partial_{i}\partial^{i} \left(\frac{\delta T}{\bar T}\right)+\frac{\dot {\bar T}}{\bar T} \partial_{i}u^{i}\right]\,.
\ee
Now we set $\zeta=0$, use the zeroth order solution $\dot {\bar T}\simeq -H \bar T$, trade temperature for density perturbations $4 \delta T/\bar T=\rgo$ and focus on scalar perturbations $u^{i}=a^{-2}u_{i}=\partial_{i}u$. The result is
\be\label{uuv}
\Delta T^{i\nu}_{;\nu}&=&-\frac{4}{3}\eta a^{-4}\partial_{j}^{2}\partial_{i}u\,,\\
\Delta N^{\mu}_{;\mu}&=&\nbz\frac{\chi_{b}\bar T}{\frac{4}{3}\rgz}  \frac{R}{(1+R)^{2}} \partial_{i}\partial^{i}\left( \frac{\rgo}{4}-H u\right)\,.
\ee
Because of \eqref{eq0} the energy conservation equation \eqref{00} is unaltered, while the momentum and number conservation in Fourier space become
\be
\frac{3}{4} \rgod+ R \rbod- \frac{q^{2}}{a^{2}} (R+1) u&=& - (R+1) \psi \,,\\
(R+1) \dot u- H u + \frac{1}{4}\rgo&=&-2 \Gamma_{\eta}u\,,\label{i1v}\\
R \left(\nbod-\frac{q^{2}}{a^{2}}u\right)&=&-R\psi+6 \Gamma_{\chi}\left(  \frac{\rgo}{4}-H u\right)\,,\label{n1v}
\ee
where, with some hindsight, we have introduce the well-known damping rates
\be
\Gamma_{\chi}=t_{\gamma}\,\frac{q^{2} }{ a^{2} } \,\frac{R^{2}}{6(1+R)^{2}}\,,\quad \Gamma_{\eta}=t_{\gamma}\,\frac{q^{2} }{ a^{2} (1+R)}  \frac{8}{45}\,. \label{Gamma}
\ee
As in the perfect fluid case, we can solve for $\nbod=\rbod$, $\dot u$ and $u$ and find a second order ordinary differential equation for $\rgo$
\be
\rgodd+\rgod \left[H \frac{1+2R}{1+R}+2 \left(\Gamma_{\chi}+\Gamma_{\eta}\right)+\mathcal{O} \left(\frac{H^{2}a^{2}}{q^{2}}\Gamma_{\chi}\right)\right]+ \label{again}\\
\quad +\rgo \frac{q^{2}}{a^{2}} c_{s}^{2}\left[1+\mathcal{O} \left(\frac{H \Gamma_{\chi} a^{2}}{q^{2}}\right)\right]=0\,.\nonumber
\ee
The two terms that we have not written down explicitly are small and can be neglected. The first is suppressed by $(aH/q)^{2}$ which is small inside the horizon, the second is suppressed by $H \Gamma_{\chi} a^{2}/q^{2}\sim H t_{\gamma}$ which is small at any time before Hydrogen recombination. The solution of this equation can be found with the WKB approximation. At leading order, the real part of the time dependent frequency is unchanged with respect to the perfect fluid case, while the imaginary part has a new contribution proportional to the sum of the dissipation rates. The solution normalized as discussed in section \ref{ss:Bpf} is
\be
\rgo(t)&=&-\frac{ 4\R^{0}}{(1+R)^{1/4}}\, \left(1-0.268 R_{\nu}\right)\,\cos \left(\int^{t} \frac{q}{a \sqrt{3}}\,dt'\right) e^{-\int^{t} (\Gamma_{\chi}+\Gamma_{\eta})dt''}\,.\label{again2}
\ee
Another convenient way to re-write the damping is by introducing the damping scale 
\be\label{qD}
q_{D}^{-2}\equiv \int \frac{dz (1+z)}{6 (1+R)  H} t_{\gamma} \left[\frac{R^{2}}{1+R} +\frac{16}{15}\right]\,,
\ee
so that\footnote{Notice that with this convention \cite{dod}, the damping scale in the power spectrum will have an additional factor of $\sqrt{2}$.} $\rgo,u\propto \exp \left(-q^{2}/q_{D}^{2}\right)$.
As we did in \eqref{usq}, we can use this result to compute $u^{i}u_{i}$. If we are interested in modes well inside the horizon we can neglect $\psi$ and terms suppressed by $aH/q$. As long as the wavelength of the modes is also much longer than the mean free path $t_{\gamma}$ we can also neglect the heat conduction term in \eqref{n1v}. Then we find
\be\label{uiui}
\ex{u^{i}u_{i}}&=& \int \frac{d^{3}q}{(2\pi)^{3}}\frac{| \R^{0} |^{2}}{(1+R)^{1/2}}\left[3 c_{s} \left(1-0.268 R_{\nu}\right)\,\sin \left(\int^{t} \frac{q}{a}c_{s}\,dt'\right) e^{-\int^{t} (\Gamma_{\chi}+\Gamma_{\eta})dt''}\right]^{2}\nonumber \\
&&\quad+\mathcal{O} \left(\frac{q^{2}}{a^{2} H^{2}},\frac{q}{a}t_{\gamma}\right)\,.
\ee
All the discussion at the end of subsection \ref{ss:Bpf} carries through, in particular \eqref{exbbp} is valid also in the presence of dissipation.


\subsubsection{Second order}\label{ss:Ds}

In principle viscous corrections can affect the dynamics of second-order perturbations which we studied in subsection \ref{ss:sB} in the perfect fluid case. In practice though, one expects viscous corrections to become relevant only at scales that are parametrically smaller than the horizon, since these are all higher-order in an expansion in derivatives. As we discussed in section \ref{intro}, all of the modes that we can realistically hope to detect, i.e.~$l\lesssim 1400$, where well outside of the horizon for the whole duration of the $\mu$-era. For these modes we can then neglect the effect of viscosity, except for taking them into account in the solution of the first-order perturbations that source the second-order evolution. Hence, the solution is again the one found in subsection \ref{ss:sB}, i.e.~\eqref{sBN}, where now for $u_{i}u^{i}$ we should use \eqref{uiui}. Focussing on $\rgt$ one finds 
\be
\ex{\rgtd}&=&-\frac{2}{3}(2+R) \partial_{t}\ex{u_{i}u^{i}}-\frac{4}{3}HR \ex{u_{i}u^{i}}\,.\label{td}
\ee
This equation tells us how the evolution of the background average energy density changes due to the dynamics of acoustic waves. There are three types of terms which source $\ex{\rgtd}$ and hence represent the backreaction of waves on the homogenous evolution: dissipation terms and oscillation terms that survive in the limit $R\rightarrow 0$ and terms of order $R$. 
\begin{itemize}
\item Dissipation terms. These are the most important and are present also for $R=0$. They arise when the time derivative in \eqref{td} hits the damping factor in \eqref{uiui}. The physical picture is that dissipation erases all waves on small scales and homogeneously redistribute their energy, hence sourcing  $\ex{\rgt}$.
\item Oscillation terms. They arise when the time derivative in \eqref{td} hits the sines in \eqref{uiui}. These terms are present also for $R=0$. The physical implication is that even if the amplitude of the waves were constant, there would be oscillations in $\ex{\rgt}$. Notice that the average energy density $\ex{T^{00}_{(2)}}$ is constant and oscillations arise just because $\rho$ is defined as the energy density in the rest frame rather than the cosmological one.
\item Terms of order $R$ are always small corrections during the $\mu$-era since $R$ increases monotonically with time reaching its maximum at the end, when $R(z_{f})\simeq 0.01$. In addition, as we will see in subsection \ref{ss:s}, all these terms should cancel in the source term for $\ex{\dot \mu}$.

\end{itemize}
We will provide more details on the dissipation terms in \ref{ss:Ds}, when we will discuss the generation of spectral distortion.


\section{During the $\mu$-era}\label{s:D}

As the universe expands, double Compton and Bremssstrahlung interactions become less and less efficient. These interactions are the leading processes that change the number of photons, since elastic Compton scattering does not create new photons. After redshift of $z_{i}\equiv 2\times 10^{6}$ double Compton and Bremssstrahlung interactions can efficiently create new photons only at very low frequencies. But Compton scattering is not fast enough to redistribute these photons to higher frequencies. The end effect is that the number of photons above a certain low-frequency threshold becomes effectively conserved, i.e.~the $\mu$-era begins. We can still describe the system using hydrodynamical equations, but the conservation of energy, momentum and baryon number needs to be supplemented by another equation for the conservation of the number of photons
\be
\left(N_{\gamma}^{\mu}\right)_{;\mu}=0\,.
\ee
 During the $\mu$-era Compton scattering is very efficient at exchanging momentum between photons and electrons (and eventually baryons). This means that perfect thermodynamic equilibrium is reached but with a conserved charge, i.e.~the number of photons. By studying the Boltzmann equation with just Compton scattering in the collision term (Kompaneets equations \cite{Kom}), one finds \cite{ref1,ref2} that the occupation number of photons is well described by the Bose-Einstein distribution with some temperature $T$ and chemical potential $\mu$. Energy momentum tensor and number density are given by
\be
T^{\mu\nu}(\vec x,t)&=&\frac{1}{a^{3}}\int \frac{d^{3}p}{(2\pi)^{3}} \frac{p^{\mu}p^{\nu}}{p^{0}} \frac{1}{e^{p^{0}/kT(\vec x,t,\hat p)+\mu(\vec x,t)}-1}\,, \label{mudef}\\
N^{\mu}(\vec x,t)&=&\frac{1}{a^{3}}\int \frac{d^{3}p}{(2\pi)^{3}} \frac{p^{\mu}}{p^{0}} \frac{1}{e^{p^{0}/kT(\vec x,t,\hat p)+\mu(\vec x,t)}-1}\,,
\ee 
where with this definition $\mu$ is dimensionless. We adopt the same sign that is commonly used in the literature, i.e.~the opposite to the one used in thermodynamics. In the tight coupling regime the mean free path is so short that an observer in the rest frame detects an \textit{isotropic} distribution, i.e.~$T$ and $\mu$ do not depend on $\hat p$, perturbations have only a monopole. Also $T$ by definition does not depend on $|p|$ and $\mu$ can found to be
\be
\mu=\mu_{\infty} e^{-2 p_{\rm peak}/p}\,,
\ee
with $p_{\rm peak}\sim 10^{-2} kT$, so that it is very close to a constant for middle and high frequencies. In the following we will simply use $\mu$ to indicate this constant value. Then the rest-frame photon number density is
\be\label{ng}
N^{0}_{\rm rest}\equiv n_{\gamma}&=&a^{-3}\int \frac{d^{3}p}{(2\pi)^{2}} \,\frac{1}{e^{p^{0}/T+\mu}-1}\nonumber\\
&=& \ngz \left[1+3 \dto+3\dtt+3\dto^{2}-\At\mu\right]\,.\label{ex2t}
\ee 
where $\At\equiv-\pi^{2}/[6\zeta(3)]$, $p^{0}p^{0}=p_{i}p_{i}a^{-2}\equiv p^{2} a^{-2}$, $\ngz\equiv \bar T^{3} \br$ with $\br\equiv 2 \zeta(3)/\pi^{2}$, and we defined the \textit{rest-frame} temperature perturbation by $\Theta_{r}\equiv \delta T/T$. In the second line we have expanded to linear order in $\mu$ and quadratic in $\Theta$. Terms quadratic in $\mu$ are very small and we neglect them. A similar computation leads to
\be\label{rgm}
T^{00}_{\rm rest}\equiv \rho_{\gamma}&=&a^{-3}\int \frac{d^{3}p}{(2\pi)^{2}} \,\frac{p^{0}}{e^{p^{0}/T+\mu}-1}\nonumber\\
&=&\rgz \left[1+4 \dto+4\dtt+6\dto^{2}-\Ao\mu \right]\,,\label{ex3}
\ee
with $\Ao\equiv 90 \zeta(3)/\pi^{4}$,  $\rgz=\ar \bar T^{4}$. As the universe keeps expanding the temperature as well as the density of electrons decreases. After redshift of about $z_{f}\equiv 5\times 10^{4}$  the interactions become so rare and the typical momentum exchanged so small that elastic Compton scattering is not efficient enough in maintaining thermodynamical equilibrium\footnote{ This is sometimes referred to as kinetic equilibrium to stress the fact that interactions isotropy photons but do not change the amplitude of their momentum.}. This time signals the end of the $\mu$-era. At any time after $z_{f}\simeq5\times 10^{4}$, any perturbation to the system cannot be efficiently thermalized. The type of distortion arising at later times is of the $y$-type and has a different frequency dependence from the distortion generated during the $\mu$-era.

 
\subsection{Inviscid fluid}\label{ss:Dzo}

In this section we will repeat the computations performed in section \ref{s:Bin} supplementing the equations of motion with an equation for the conservation of the photon number. The main result is that $\mu$ is sourced only by terms of order $\nbz/\ngz$, which was estimated in \eqref{r}. Let us to start considering the idealized case in which photons behave as a perfect fluid. We can write the photon number current as
\be
N^{\mu}_{\gamma} \equiv u^{\mu} n_{\gamma}\,,
\ee
where $n_{\gamma}$ is the photon number density in the rest frame of the fluid given in \eqref{ng}. For an imperfect fluid, this expression is corrected by a heat conduction term. This will be discussed in section \ref{Dv} and appendix \ref{a:chi} and turns out to be a negligible effect.

We study first the evolution of the zeroth-order homogeneous background. As long as we neglect viscous corrections, the entropy is conserved  and the system evolves along an adiabat. This is true both at the homogeneous and inhomogeneous level, therefore the zeroth order results will be sufficient to understand the behavior at higher orders in perturbation theory. The number of photons is conserved in a comoving volume
\be
\ngzd+3H\ngz=0\,.
\ee
Adding this to the equations of subsection \ref{Bv}, and using \eqref{ng} and \eqref{rgm} to rewrite $\ngg$ and $\rho_{\gamma}$ in terms of $\delta T$ and $\mu$, on finds
\be
\nbzd&=&-3H\nbz\,,\\
\Tzd&=&- H \Tz \left[1+\frac{3}{8} r \frac{\pi^{6}}{\pi^{6}-405 \zeta(3)^{2}}+\mathcal{O} \left(r^{2}\right)\right]\nonumber\\
&\simeq &- H \Tz \left(1+\frac{3}{8} \,r\, 2.5\right)\,,\\
\dot {\bar \mu}&=& -\frac{27}{4}r H \frac{\pi^{4}\zeta(3)}{\pi^{6}-405 \zeta(3)^{2}}+\mathcal{O} \left(r^{2}\right)\,,\nonumber\\
&\simeq &-2.1 \,r H\,.
\ee
So we find that a homogeneous \textit{negative} $\mu$-distortion is created by the adiabatic expansion. This effect to was first noticed in \cite{BE} and it is sometimes referred to as production of $\mu$-distortion by \textit{adiabatic cooling} of electrons.  Intuitively, electrons and baryons would tend to cool down faster than the photons. Because of the continuous interactions, the electrons and baryons extract energy out of the photons. This results in a negative $\mu$-distortion. As the name suggests, no entropy is created in this process. Notice also that the amount by which the temperature evolution deviates from $a^{-1}$ is different from what we found before the $\mu$-era \eqref{Tbm}. The typical amount of $\mu$-distortion produced by this effect is of order the baryon-to-photon ratio $r$. Numerically integrating over the whole $\mu$-era, we estimate the effect of adiabatic cooling as 
\be
\bar \mu=-\int_{z_{f}}^{\infty}  \frac{dz}{(z+1)} e^{- \left(\frac{z}{z_{i}}\right)^{5/2}} \,2.1 \times r  \simeq -1.6\times 10^{-9}\,,
\ee
where rather than a sharp cutoff at high redshift we use the exponential suppression at $z_{i}$ obtained from the analytical solution of \cite{HSS}. Notice that taking into account baryon temperature is essential in order to capture this effect. As we mentioned before, baryon temperature is a small effect and it is typically negligible unless all other effects cancel precisely. This is exactly what happens here: neglecting baryon temperature $T\propto a^{-1}$, and so $n_{\gamma}\propto a^{-3}$. Then the conservation of photon number is automatically satisfied by the solution even without enforcing $N^{\mu}_{\gamma;\mu}=0$ and no $\mu$-distortion is created. At first and second order waves propagate on top of the homogeneous background. Neglecting dissipation, each fluid element expands and contracts adiabatically, moving along the same adiabat as the background. Therefore a discussion analogous to the one above holds. Neglecting baryon temperature (or equivalently terms of order $\nbz/\ngz$) no spectral distortion is created. In order to understand this from a slightly different point of view, let us discuss more in detail the conservation of entropy. 

 
\subsection{Entropy considerations}\label{ss:s}

In this subsection we use again the argument presented in the introduction \eqref{ss:s}. Neglecting viscous corrections, entropy is covariantly conserved. We can therefore define\footnote{A definition of the entropy density current that accounts for dissipative effect is in given in \eqref{smu}} an entropy-density current $s^{\mu}\equiv s u^{\mu}$, where $s=-u_{\mu}s^{\mu}$ is the rest-frame entropy density. This is conserved during the adiabatic evolution and satisfies $s^{\mu}_{;\mu}=0$. Both the entropy of baryons and photons is proportional to their number density\footnote{This can be checked using $T s_{x}=\rho_{x}+p_{x}-\mth^{x}n_{x}$. The baryon chemical potential is given in \eqref{mub}.}, hence up to terms of order $\nbz/\ngz$, discussed in the previous subsection, we can approximate $s^{\mu}\simeq s_{\gamma}^{\mu}$. Using $T s_{\gamma}=\rho_{\gamma}+p_{\gamma}-\mth n_{\gamma}$ one finds
\be
s_{\gamma}=\frac{4}{3}\ar \bar T^{3}\left(1+3 \dto+3\dtt+3\dto^{2}+u^{0}-\Atr\mu\right)\,,\label{As}
\ee
where $\Atr\equiv 135\zeta(3)/(2\pi^{4})$. Combining this expression with \eqref{ex2t} one finds
\be
\partial_{t} \left(\frac{s_{\gamma}}{n_{\gamma}}\right)=\dot \mu 
\frac{4}{3}\frac{\ar}{\br} \left(\At-\Atr \right)\simeq 1.9\, \dot \mu\,.\label{mudsum}
\ee
Using the conservation of entropy and photon number $s^{\mu}_{;\mu}=n^{\mu}_{;\mu}=0$, we then find $\dot\mu=0$, which result is valid up to correction of order $\nbz/\ngz$. This result agrees with the discussion in the last subsection. As we will now see, once entropy is not conserved, things become more interesting.

%
%
 
\subsection{Viscous fluid}\label{Dv}

Let us include the effect of dissipation during the $\mu$-era. The expressions \eqref{deltaN} for the dissipative corrections obtained in subsection \ref{Bvf} are still valid, but we need to compute the heat conduction for the photon number conservation $\chi_{b}$. We leave the details of the derivation to appendix \ref{a:chi}, and describes shortly the result. In the presence of dissipation, the velocity of energy transport differs from the one of number transport. Physically this can be understood as follows. Imagine two regions of the fluid with a different temperature. Suppose the same number of particles diffuse from the hotter to the colder region and viceversa. Then there is not net number transport, but there is a non-vanishing energy transport since particles coming from the hotter region carry in average more energy. There is another way of thinking about heat conduction. Consider a (very weakly interacting) gas of photons in a box. Consider the initial configurations in which the left hand side of the box has a certain temperature and number density different from those in the the right-hand side of the box. Imagine to choose temperatures and number densities such that energy density and pressure are exactly the same on the two sides of the box. For an ideal fluid, this configuration does not evolve. Conversely, once heat conduction is taken into account, a heat and number flow turns on and evolves the system towards equilibrium.

The final result (see appendix \ref{a:chi}) for the heat conduction of photons is
\be
\chi_{\gamma}T=\frac{4}{3} \rgz t_{\gamma}\,\frac{2\pi^{4}}{45\zeta(3)}\frac{R^{2}}{\bar\mu}\,,
\ee
where the apparent divergence as $\bar \mu\rightarrow 0$ is fictitious since $\bar \mu$ cancels in the expression for $\Delta N^{\mu}$.


\subsubsection{Zeroth order}

At zeroth order in perturbations, the only new effect with respect to the perfect fluid case of subsection \ref{ss:Dzo} is given by the bulk viscosity, which is the only non-vanishing viscous coefficient at this order. Some distortion proportional to $\zeta$ is generated, but as we discussed, this effect is very small, of order $r^{2}$. In passing, this means that the background evolution is very close to adiabatic, i.e.~no entropy is generated. In the following, we will neglect bulk viscosity. Since the background is still homogenous and isotropic, the relations \eqref{did} are still valid.

 
\subsubsection{First order}

During the $\mu$-era, there is an extra equation in addition to those discussed in subsections \ref{ss:Bpf} and \ref{ss:Bfv} $N_{\gamma;\mu}^{\mu}+\Delta N_{\gamma;\mu}^{\mu}=0$. Using the results of appendix \ref{a:chi} and \eqref{ex2t} and \eqref{ex3} one finds
\be
N^{0}&=&\ngz\left[\frac{3}{4} \rgo+ \frac{405 \zeta(3)^{2}-\pi^{6}}{6 \pi^{4}\zeta(3)}\mu\right]\,,\\
N^{i}&=&\ngz u^{i}\,\\
\Delta N^{i}&=& \ngz \frac{R^{2}}{(1+R)^{2}}t_{\gamma}  \left[\partial^{i}\dto+\partial^{i}\dm-u^{i}H\right]\,.
\ee
The resulting conservation equation is
\be
-4 A \dmd+4 \frac{q^{2}}{a^{2}} u -3 \rgod=- \frac{q^{2}}{a^{2}}\frac{R^2 \,t_{\gamma}}{(R+1)^2 } \,\left[-4 H u+2 \left(\frac{45 \zeta (3)}{\pi ^4}+2\right) \dm+\rgo \right]\,.
\ee
For modes well \textit{inside} the horizon, as we did previously, we can neglect gravity perturbations, solve for $u$, $\dot u$ and $\rbod$ and obtain a system of a one first- and one second-order differential equations for $\dm$ and $\rgo$. Notice thought that the longest mode inside the horizon during the $\mu$-era is $l\sim1400$ entering around $z_{f}\simeq 5\times 10^{4}$. We will discuss modes \textit{outside} of the horizon shortly. By keeping only terms that are at most linear in either $\dm$ or $t_{\gamma}$, inside the horizon one finds
\be
\dmd&=&-\frac{3 \pi ^4 \zeta (3) }{\left(\pi ^6-405 \zeta
   (3)^2\right) } \frac{R^2 t_{\gamma} }{(R+1)^2}\left(\frac{q^{2}}{a^{2}}
   \rgo-3 H\rgod\right)\\
   &=& 0.93\times  \frac{R^2 t_{\gamma} }{(R+1)^2}\left(\frac{q^{2}}{a^{2}}
   \rgo-3 H\rgod\right)\,,
\ee
and for $\rgo$ the same equation as before the $\mu$-era, \eqref{again}, up to terms of order $t_{\gamma}H$ and $q/(aH)$. Since the solution for $\rgo$ is again \eqref{again2} at leading order, the above equation tells us how the oscillations of the fluid source short scale ($l\gtrsim 1400$) perturbations in $\mu$ at linear order. 

Long perturbations in $\mu$, i.e.~$l\ll 1400$, are of more direct observational interest. Since these modes were well outside the horizon during the $\mu$-era, we can compute their evolution by neglecting derivatives. Since all viscous corrections (apart from the bulk viscosity which is extremely small) come with a higher number of derivatives, they are all negligible. Then the dynamics is the same adiabatic evolution that we studied in subsection \ref{ss:Dzo}. Up to terms suppressed by the photon to baryon ratio, $\dmd=0$.

 
\subsubsection{Second order}\label{ss:Dsv}

One way to derive an equation for the generation of $\mu$-distortion at second order was presented in the introduction, subsection \ref{ss:sum}, using the non-conservation of the entropy caused by viscous corrections. Here we present an alternative derivation that does not make use of any entropy consideration. 

As discussed at the end of subsection \ref{ss:Ds}, the leading sources for the second order perturbations arise when the time derivatives hit the dissipation factors in $\ex{u_{i}u^{i}}$. Terms of order $R$ are, on the one hand, small corrections during the $\mu$-era and can be neglected. On the other hand, according to the discussion of subsection \ref{ss:s}, we know that up to viscous corrections, the system evolves adiabatically and $\ex{\dot\mu}=0$. The terms of order $R$ in \eqref{td} are not viscous correction, e.g.~since they survive as $t_{\gamma}\rightarrow0$, and therefore can not source $\ex{\dot\mu}$. We verified this explicitly to leading and first subleading\footnote{In order to have a cancellation at this order one should include in \eqref{wont} a term subleading in $aH/q$, coming from \eqref{usq} when the time derivative on the right hand side acts on the amplitude of $\rgo$ rather than on the cosine.} order in $aH/q$, where $q$ is the wave number of the short-scale dissipating perturbations. This is important since it ensures that all the energy stored in primordial perturbations is conserved during their evolution and is available at the time the perturbations reach the dissipation scale. From now on we set $R=0$ and neglect baryon conservation.

Instead of taking the average as we did in subsection \ref{ss:Ds} and \ref{ss:sB}, we work with the full energy and photon number conservation equations at second order
\be
\left(4H+\partial_{t}\right)T^{00}_{(2)}+\partial_{i}T^{0i}_{(2)}&=&0\,,\label{T2}\\
\left(3H+\partial_{t}\right)N^{0}_{(2)}+\partial_{i}N^{i}_{(2)}&=&0\,,\label{N2}
\ee
where the viscous corrections in these formulae can be omitted since they are higher order in derivatives. On the other hand, as we will see shortly, it is essential to account for the dissipation damping in the solution of the first order perturbations. To simplify the above equations we used $a^{2}H T^{ii}_{(2)}=H T^{00}_{(2)}$. Given that
\be
T^{00}_{(2)}&=& \frac{4}{3}\rgz \left[\frac{3}{4}\rgt+ u_{i}u^{i}\right]\,,\\
T^{0i}_{(2)}&=& \rgz \left(\rgo u^{i}+u^{i}_{(2)}\right)\,,\\
N^{0}_{\gamma(2)}&=& \ngz \left[\ngt+ \frac{1}{2} u_{i}u^{i}\right]\,,\\
N^{i}_{\gamma(2)}&=& \ngz \left(\ngo u^{i}+u^{i}_{(2)}\right)= \ngz \left(\frac{3}{4}\ngo u^{i}+u^{i}_{(2)}\right)\,,
\ee
we can solve \eqref{T2} for $u^{1}_{(2)}$ and substitute it into \eqref{N2}. The result is
\be
\partial_{t}\left(4 \ngt-3\rgt\right)=\partial_{i} \left(\rgo u^{i}\right)+2 \partial_{t} \left(u_{i}u^{i}\right)\,.
\ee
Using \eqref{ex2t} and \eqref{ex3}, we see that this is an equation just for $\mu$:
\be\label{mufin}
\boxed{
\dot \mu= \left(3\Ao-4\At \right)^{-1} \left[2 \partial_{t}\left(u_{i}u^{i}+3\tru^{2}\right)+\partial_{i} \left(\rgo u^{i}\right)\right]\,.}
\ee
As a check of this equation, let us take its expectation value, hence dropping the total derivative:
\be
\ex{\dot \mu}&=&-
\frac{2}{4\At-3\Ao}\partial_{t} \left(\ex{u^{i}u_{i}+3\tru^{2} }\right)\nonumber\\
&\simeq& -0.93 \,\partial_{t} \left(\ex{u^{i}u_{i}+3\tru^{2} }\right)\nonumber \,,\\
\ee
Using the relation between first order quantities $\tru=\frac{\rgo}{4}$, we can rewrite \eqref{exbbp} as
\be
\ex{u^{i}u_{i}}+3 \ex{\tru^{2}} =\frac{3}{8}\ex{\ex{\rgo^{2}}}_{p}\,.
\ee
The equation for $\mu$ simplifies to
\be
 \partial_{t}\ex{ \mu}&=&- \frac{9 \pi^{4}\zeta(3)}{2 \left[\pi^{6}-405 \zeta(3)^{2}\right]} \frac{1}{4}\partial_{t} \ex{\ex{\rgo^{2}}}_{p}\nonumber\\
 &\simeq& -1.40\times \frac{1}{4}\partial_{t} \ex{\ex{\rgo^{2}}}_{p}\,,\label{qf}
 \ee
This result agrees with the derivation in subsection \ref{ss:sum} and can be interpreted as
\be
 \partial_{t}\ex{ \mu}&=&-1.40\times \frac{c_{s}^{2}}{1+w} \partial_{t} \ex{\ex{\rgo^{2}}}_{p}\,,
 \ee
where we see that the origin of the factor of $3/4$ found in \cite{2t2} is the relativistic correction $(1+w)^{-1}$ to the energy of wave, which we discuss in appendix \ref{a:cs}.
For modes of observational interest one can drop the total derivative term in \eqref{mufin}. Then integrating over time we find
\be\label{mu0f}
\mu_{0}(\vec x,t_{f})= \frac{2}{3\Ao-4\At}  \left[u_{i}u^{i}+3\tru^{2}\right]^{f}_{i}\,,
\ee
where the index ``$0$'' is a reminder that this is a monopole (the same in every direction) and $ \left[\cdot\right]^{f}_{i}$ indicates the difference of its argument between the beginning of the $\mu$-era at $z_{i}\simeq 2\times 10^{6}$ and the end $z_{f}\simeq 5\times 10^{4}$. Using \eqref{mu0f} one can compute the momentum space power spectrum at the end of the $\mu$-era (details are given in appendix \ref{a:ps}, see \eqref{Pmu}) in the limit of small momenta
\be
P_{\mu_{0}}(q,\tau_{f})&=& \frac{2\left(1-0.268 R_{\nu}\right)^{2}}{3 \left(3\Ao-4\At\right)} \int \frac{d^{3}k}{4\pi} \frac{2\pi^{2}\Delta_{\R}^{4}(k)}{k^{6}}  \left\{ \left[ e^{-2k^{2}/q_{D}^{2}}\right]^{f}_{i}\right\}^{2}\\
&\simeq&2\times 10^{-14}\,  \frac{2\left(1-0.268 R_{\nu}\right)^{2}}{3 \left(3\Ao-4\At\right)} 2\pi^{2}\Delta_{\R}^{4}(k_{p})\\
&\simeq&\frac{4}{q_{D}(z_{f})^{3}}\, \frac{2\left(1-0.268 R_{\nu}\right)^{2}}{3 \left(3\Ao-4\At\right)} 2\pi^{2}\Delta_{\R}^{4}(k_{p})
\ee

%
%
%
%
%

 
\section{After the $\mu$-era}\label{s:A}

As the universe expands, after $z_{f}=5\times 10^{4}$, kinetic equilibrium is lost since the momentum exchanged between photons and electrons is of order $T/m_{e}\ll1$. From this moment onword, thermodynamic equilibrium is lost and no additional $\mu$-distortion can be generated. In this section, we provide a transfer function to relate the $\mu$-distortion at the end of the $\mu$-era to the one we can measure at late times\footnote{In \cite{Ganc:2012ae} this transfer function was taken to be just the projection of the monopole at the end of the $\mu$-era onto higher multipoles in the late time CMB sky. As we show here, this is a very good approximation for large scales, $l\ll 1000$.} in the CMB. It turns out that the Boltzmann equation is particularly simple, so we abandon the hydrodynamic equations for this section and switch to the kinetic description. 

We consider the Boltzmann equation at linear order in perturbations, neglecting double Compton scattering and Bremssstrahlung in the collision term. We make a Bose-Einstein ansatz for the photon occupation number, with constant $\mu$, which is known to be a good description for not so low frequencies $\nu/T\gg 10^{-2}$. We neglect polarization and multiples above the dipole, since these are suppressed in the tight coupling regime. In general one has $\mu(t,\vec x,\hat p,p)$ and our conventions for the decomposition in spherical moments are reviewed in appendix \ref{a:kin}. Going to Fourier space and introducing the cosine between the wavenumber $\vec q$ and the photon momentum (or direction of observation) $\hat p$, $\chi\equiv \hat p \cdot \hat k $, we find
\be\label{Bmu}
\mu'+ i q \chi \mu= \frac{a}{t_{\gamma}} \left(\mu_{0}-\mu\right)\,,
\ee
where a prime denotes derivatives with respect to conformal time $d\tau\equiv adt$. Notice that this is very similar to the analogous equation for temperature perturbations, with the major difference being that there is no velocity sourcing a dipole for $\mu$. Following this analogy, we solve \eqref{Bmu} in two steps. First we use the tight coupling expansion to find a solution in the epoch between the end of the $\mu$-era and the last scattering surface, when the photon mean free path $t_{\gamma}$ is the shortest distance in the problem. This gives us some $\mu$ monopole $\mu_{0}$ and dipole $\mu_{1}$ at the last scattering surface. Second, using the integral along the line of sight we solve \eqref{Bmu} in the epoch between last scattering and arrival on the earth in the present day. This free streaming evolution relates $\mu_{0}$ and $\mu_{1}$ to the observed multipole at late times.

 
\subsection{Tight coupling} 

Projecting on the zeroth and first Legendre polynomials, we derive the coupled equations for the $\mu$ monopole $\mu_{0}$ and dipole $\mu_{1}$
\be
\mu_{0}'+ q \mu_{1}&=&0\,,\\
\mu_{1}'- \frac{q}{3}\mu_{0}&=&-\frac{a}{t_{\gamma}}\mu_{1}\,.
\ee
As usual, by projecting on higher Legendre polynomials, one can verify that in the tight coupling limit $t_{\gamma}\rightarrow 0$, the higher multiples are suppressed by powers of $q t_{\gamma}$ and hence can be neglected. We algebraically solve the above equations for $\mu_{1}$ and $\mu_{1}'$ and plug the solution back into the time derivative of the first equation. We find
\be
\mu_{0}''+\mu_{0}' \frac{a}{t_{\gamma}}+\frac{q^{2}}{3}\mu_{0}=0\,.
\ee
The two WKB solutions can be expanded in $qt_{\gamma}\ll1$. The one that decays more slowly is
\be
\mu_{0}(q,t)&=&\mu_{0}(q,t_{f}) \exp \left({-\int_{t_{f}}^{t} \frac{q^{2}}{a^{2}} t_{\gamma}dt'}\right) \label{dd}\\
&\equiv &\mu_{0}(q,t_{f}) \exp \left({-\int_{t_{f}}^{t} \Gamma_{\mu} dt'}\right) \equiv \mu_{0}(q,t_{f})\, e^{-q^{2} \left[\qmd(t)^{-2}-\qmd(t_{f})^{-2}\right]} \,,\nonumber
\ee
where $\mu_{0}(q,t_{f})$ is the value of the $\mu_{0}$ monopole at the end of the $\mu$-era, \eqref{mu0f}, and we introduced the decay rate $\Gamma_{\mu}$ and the $\mu$-dissipation scale $\qmd$, both functions of time. It is useful to estimate $\qmd$ since it is the smallest scale on which we can observe some inhomogeneity in $\mu$. We can use
\be
t_{\gamma}^{-1}&=&\sigma_{T} \left(1-\frac{Y}{2}\right) \left(1+z\right)^{-3}\frac{\Omega_{b} \rho_{\rm crit}}{m_{b}}\\
&=&4.5 \times 10^{7}\, {\rm Mpc}^{-1}  \left(1+z\right)^{3}\,,
\ee
where the Helium fraction $Y=0.23$ enters because of the relation between electron and baryon number density $n_{e}= \left(1+Y/2\right) n_{b}$, $\sigma_{T}\simeq 66\, {\rm fm}^{2}$ is the Thomson cross section, $H_{0}=2.4 \times 10^{-4} \rm{Mpc}^{-1}$ is the Hubble constant nowadays from it follows that today's critical density is $\rho_{\rm crit}\simeq (2.5 \times 10^{-3} {\rm eV})^{4}$. Then\footnote{The integral is supported at late times, so the upper bound of the integral is irrelevant as long as it is much larger than the redshift at last scattering $z_{LLS}\simeq 1100$}
\be
\qmd&=& \left[ \int_{1100}^{+\infty} t_{\gamma} \frac{dz (1+z)}{H}\right]^{-1/2}\label{qDm}\\
&\simeq& 0.084 \times {\rm Mpc}^{-1}\,,
\ee
corresponding to a multiple $l_{\mu D}\simeq 1200$, which is comparable with the damping scale of temperature anisotropies.

 
\subsection{Free streaming}\label{Afs}

As free electrons combine with protons to form neutral hydrogen around $z\simeq 1100$, the mean free path of photons grows larger and larger and the tight coupling approximation breaks down. In fact, most of the photons we observe in the CMB last scattered very close to $z\simeq 1100$, i.e.~have been free steaming since recombination. It is well known how to evolve temperature anisotropies during free streaming, and very similar techniques can be used to solve \eqref{Bmu}. First we rewrite it as
\be
\frac{d}{d\tau} \left[\mu(\tau,q,\chi) \exp \left(iq\chi \tau+\int^{\tau} d\tilde\tau \frac{a}{t_{\gamma}}\right) \right]=\frac{a}{t_{\gamma}}\mu_{0}(\tau,q) \exp \left(-iq\chi \tau-\int^{\tau} d\tilde\tau  \frac{a}{t_{\gamma}}\right) \,.
\ee
The line-of-sight solution is 
\be
\mu(\tau,q)=\int^{\tau}_{\tau_{0}} d\tilde \tau g(\tilde \tau,\tau) \mu_{0}(\tilde \tau,q) e^{iq\chi (\tilde\tau-\tau)}\,,
\ee
where $\tau_{0}$ is some very early time and we have defined the visibility function (the same as for temperature anisotropies)
\be
g(\tilde \tau,\tau)\equiv \frac{a(\tilde \tau)}{t_{\gamma} (\tilde \tau)} \exp \left(-\int^{\tau}_{\tilde \tau} \frac{a(\tau')}{t_{\gamma} (\tau')} d\tau'\right)\,.
\ee
We can project on the various multipoles by multiplying this solution times $i^{l} P_{l}(\chi)$ and integrating over $\chi$. Using the identity
\be
i^{l}\int^{1}_{-1}\frac{d\chi}{2} P_{l}(\chi) e^{iq\chi (\tilde\tau-\tau)}=j_{l}\left[ q \left(\tau-\tilde \tau\right)\right]\,,
\ee
we find
\be
\mu_{l}(\tau,q)&=&\int^{\tau} d\tilde \tau g(\tilde \tau,\tau) \mu_{0}(\tilde\tau,q)j_{l}\left[ q \left(\tau-\tilde \tau\right)\right] \,,\label{ft}\\
&=&\mu_{0}(q,\tau_{f})M_{l} (q,\tau)\,,
\ee
where in the second line we used the result of the tight coupling analysis \eqref{dd} and introduced the transfer function
\be
M_{l}(q,\tau)\equiv\int^{\tau}_{\tau_{f}} d\tilde \tau g(\tilde \tau,\tau) \exp \left({-\int_{\tau_{f}}^{\tau} \frac{q^{2}}{a^{2}} at_{\gamma}d\tau'}\right) j_{l}\left[ q \left(\tau-\tilde \tau\right)\right]
\ee
This integral can in principle be computed numerically with the exact visibility function. Analogously to what happens for temperature perturbations, we can try to consider small $l$, say $l\ll 1000$ and approximate $g$ as a delta function around the last scattering surface $\tau_{LSS}$, where $\mu_{0}$ and $j_{l}$ vary slowly. Then
\be
\mu_{l}(\tau,q) \simeq  \mu_{0}( \tau_{f},q)j_{l}\left[ q \left(\tau- \tau_{LSS}\right)\right] \qquad \text{(tentative)}.\label{proj}
\ee
This is the transfer function used in \cite{Ganc:2012ae} and it is an excellent approximation for large scales for the goal of computing $\mu T$ correlations. For $\mu\mu$ correlations, things are more complicated. In our formalism the late time angular power spectrum is
\be
C_{l}^{\mu\mu}=\int \frac{d^{3}q}{(2\pi)^{3}} P_{\mu_{0}}(\tau_{f},q) |M_{l}(q,\tau)|^{2}\,.
\ee
where $ P_{\mu_{0}}(\tau_{f},q)$ is the momentum-space three-dimensional power spectrum of $\mu_{0}$ at the end of the $\mu$-era. The transfer function $M_{l}(q,\tau_{\rm late})$ is effectively zero for modes $q\gg \qmd(z_{LLS})$, with $z_{LSS}\simeq 1100$ the redshift of the last scattering surface. Hence we are only interested in the $q\rightarrow 0$ limit of $ P_{\mu_{0}}$, which is computed in appendix \ref{a:ps} and given in \eqref{Pmu}. In this approximation
\be
C_{l}^{\mu\mu}=P_{\mu_{0}}(\tau_{f})\int \frac{d^{3}q}{(2\pi)^{3}}  |M_{l}(q,\tau)|^{2}\,.
\ee
If we approximated the visibility function as a delta function, we would get
\be
C_{l}^{\mu\mu}=P_{\mu_{0}}(\tau_{f})\int \frac{d^{3}q}{(2\pi)^{3}}  j_{l}(qr_{L})^{2}e^{-2q^{2}/\qmd(z_{LSS})^{2}} \,.
\ee
This integral is supported on large values of $q$, and cut off by the exponential diffusion damping. But the thickness of the last scattering surface suppresses the small scale power spectrum by an amount comparable with the diffusion damping and so it can never be neglected, not even on large scales. An analytical expression can nevertheless be found by employing the flat sky approximation and approximating the visibility function as a Gaussian of conformal-time width $\sigma_{\tau}^{2}$. Then one finds
\be
C_{l}^{\mu\mu}\propto \frac{\Delta_{\R}(k_{p})^{4}}{q_{D}(\tau_{f})^{3}} \frac{\widetilde{\qmd}(\tau_{LSS})}{r_{L}^{2}} e^{-l^{2}/ l_{\mu D}^{2}}\,,
\ee
where $l_{\mu D}\sim r_{L} \qmd$ and $\widetilde{\qmd}^{-2}\equiv \qmd^{-2}+\sigma_{\tau}^{2}$. A few comments on this result are in order. The small scale pertubations present in two patches of the sky, separated by a large angular scale, $l\ll1000$, are independent random variables, so their correlation is just white noise. In fact, the $\mu$ power spectrum is $l$ independent for small $l$, as expected for white noise.

 
\section*{Acknowledgments}

It is a pleasure to thank Jens Chluba for useful discussions. E.~P.~is supported in part by the Department of Energy grant DE-FG02-91ER-40671. M.~Z.~is supported in part by the National Science Foundation grants PHY-0855425,
AST-0506556 and AST-0907969, and by the David \& Lucile Packard and the John D. \& Catherine
T. MacArthur Foundations.


\appendix

 
\section{Conversions between single fluid and two fluid descriptions}

The photon-baryon-electron plasma froms a single fluid with $\rho=\rho_{\gamma}+\rho_{b}$ and $p= p_{\gamma}=\rho_{\gamma}/3$. We define\footnote{In principle the derivative defining the speed of sound should be performed along an adiabatic transformation, but since all the processes we consider are reversible, this is the same as constant total entropy $S$.}
\be
w\equiv \frac{p}{\rho}=\frac{1}{3+4R} \,,\quad c_{s}^{2}\equiv \left(\frac{\partial p}{\partial \rho}\right)_{S,N_{b}}=\frac{1}{3(1+R)}\label{wsol}
\ee
where $S$ is the total entropy and $N_{b}$ the total number of baryons. The fastest way to compute $c_{s}^{2}$ is to notice that the background evolves adiabatically conserving the baryon number. Therefore one can use the chain rule to write $c_{s}^{2}=\dot p/\dot \rho$. This gives the correct result. To make more transparent the role played by the conservation of entropy and baryon number, in the following we provide an alternative (longer) derivation.

There are six relevant thermodynamical variables: $S,\,N_{b},\,p,\,V,\,T$ and $U$. By the definition of $c_{s}^{2}$ we are instructed to keep the first two fixed and take a derivative in the direction $\rho\equiv U/V$. To compute this derivative we need other three independent equations that relate the extra three variables. The first equation is given by the integrated form of the second law of thermodynamics
\be\label{sL}
TS=U+pV-\mu_{b} N_{b}\,.
\ee
A second an third equations are given by explicit expression for the pressure $p$ and entropy density $s\equiv S/U$. Neglecting baryon temperature, one has
\be
s=\frac{4}{3} \frac{\pi^{2}}{15} T^{3}\,,\quad p=\frac{1}{3} \left(\rho-\mu_{b} n_{b}\right)\,,\label{st}
\ee
where we have used the energy density $\rho\equiv U/V$ and baryon number density $n_{b}\equiv N_{b}/V$. The equation for $s$ comes just from the black body formula, while the one for $p$ uses $\rho=\rho_{\gamma}+\rho_{b}$ and from \eqref{mub} $\rho_{b}\simeq m_{b} n_{b}\simeq \mu_{b} n_{b}$. Notice that this relation implies that the photon-electron-baryon plasma is \textit{not} a barotropic fluid, i.e.~$p\neq p(\rho)$. Using these three equations one finds
\be
 \left(\partial_{\rho} \log s\right)_{S,N}= \left(\partial_{\rho} \log n_{b}\right)_{S,N}=- \left(\partial_{\rho} \log V\right)_{S,N}\,,
\ee
which can be solved for $\left(\partial T/\partial \rho \right)_{S,N_{b}}$. Solving \eqref{st} and \eqref{sL} for $n_{b}$ as function of $\rho$ and $T$ and computing the appropriate derivatives, one derives the well-known result
\be
c_{s}^{2}\equiv \left(\frac{\partial p}{\partial \rho}\right)_{S,N_{b}} =\frac{1}{3(1+R)}\,.
\ee
where as usual we have defined the baryon loading $R\equiv 3 \rho_{b}/(4 \rho_{\gamma})$ for which $\dot R=H R$.

%
%
Some useful relations are
\be
w&=&\frac{3 c_{s}^{2}}{4-3c_{s}^{2}}\,,\quad c_{s}^{2}=\frac{4 w}{3 (1+w)}\,,\\
\frac{\dot w}{w^{2}}&=&-4 HR\,,\quad \frac{\partial_{t} c_{s}^{2}}{H c_{s}^{2}}=-\frac{R}{1+R}\,.
\ee


\section{Initial conditions}\label{a:R}

In this appendix we derive the initial conditions for the amplitude of the pressure waves in the fluid. We are interested in radiation domination so we neglect dark matter. We also neglect neutrinos which will be discussed elsewhere. Their effect is about $10\%$. So we are left with a single photon-baryon-electron fluid (in the tight coupling regime) coupled to gravity. Following \cite{Boubekeur:2008kn} we perform the computation using a $P(X)$ Lagrangian which is equivalent to a single perfect fluid. We work in comoving gauge, i.e.~$u=0$ for $u$ the velocity potential and neglect tensor and vector perturbations. Notice that in this gauge $\delta \rho\neq0$. After solving the constraints from gravity one gets the simple second order action
\be
S_{2}=\int dt d^{3}x a^{3}\frac{\epsilon}{c_{s}^{2}} \left[\dot \R^{2}-\frac{c_{s}^{2}}{a^{2}} \left(\partial \R\right)^{2}\right]\,,\label{Ract}
\ee
where $\R$ stands for curvature perturbations on comoving slices\footnote{This quantity is called $\zeta$ in \cite{Maldacena}}. Its gauge invariant definition is
\be\label{def}
\R\equiv \frac{A}{2}+H \delta u\,,
\ee
where as defined\footnote{This is the notation of \cite{Weinberg}.} in \eqref{mper}, $\delta g_{ij}=a^{2} \left(A \delta_{ij}+\partial_{i}\partial_{j}B\right)$. During radiation domination $\epsilon=2$ and for the photon-baryon-electron plasma $c_{s}^{-2}=3 \left(1+R\right)$, with the baryon loading $R\equiv 3\bar \rho_{b}/(4\bar\rho_{\gamma})$. Much before matter-radiation equality $R\ll1$ so we neglect it in the following.

The equation of motion for $\R$ in Fourier space is
\be\label{Reom}
\R''+2aH \R'+q^{2} c_{s}^{2}\R=0\,,
\ee 
with the conformal time $\tau\equiv \int dt/a=1/(aH)=2t/a$. Notice that $\tau>0$ and runs from zero to positive infinity. The solution of \eqref{Reom} asymptoting a constant $\R^{(0)}$ in the far past ($\tau,a\rightarrow 0$) is 
\be
\R&=&\R^{(0)}\frac{aH}{q c_{s}}\sin \left(\frac{q}{aH} c_{s}\right)\\
&=&\R^{(0)}\frac{a}{q c_{s}} \frac{1}{2t}\sin \left(\int \frac{q}{a} c_{s}\,dt\right)\,.
\ee
Notice that with the definition \eqref{def}, this result is valid in any gauge, both inside and outside of the horizon, during radiation domination. If we consider modes well inside the horizon $q/(aH)\gg1$, synchronous and Newtonian gauges agree. In either of these two gauges, the gravitational potential is very small on scales smaller than the horizon so we have $\R\simeq H\delta u$. Again neglecting gravity perturbations, the conservation equation for the fluid is
\be
\dot \ru = (1+w) \frac{q^{2}}{a^{2}} \delta u\,.
\ee
Integrating we find
\be
\ru&=&- \frac{1+w}{c_{s}^{2}}\,\R^{0}\,\cos \left(\int \frac{q}{a} c_{s}\,dt\right)\\
&=&- 4\R^{0}\,\cos \left(\int \frac{q}{a} c_{s}\,dt\right)\,,\\
\ee
which agrees with  (6.4.11) of \cite{Weinberg} and (121) of \cite{Bashinsky}.

 
\section{Derivation of the viscous parameters}\label{a:s}

For some conserved charges $N_{A}$ with chemical potentials $\mth^{A}$, the second law (sum over repeated index $A=0,\dots,q$)
\be\label{mth}
TdS=dU+pdV-\mth^{A} dN_{A}\,,
\ee
can be rewritten in terms of densities $s\equiv S/V$, $\rho\equiv U/V$ and $n_{A}\equiv N_{A}/V$. Then
\be
T ds=d\rho-\mth^{A} dn_{A}+\frac{dV}{V}\left(\rho+p-\mth^{A} n_{A}-Ts\right)\,.
\ee
The volume is so far unspecified. It is convenient to define the volume such that it contains a fixed number of say $N_{0}$ particles. Then $dN_{0}=0$ and hence $V dn_{0}=-n_{0} dV$. We introduce the entropy per particle of type ``$0$'', $\sigma\equiv s/n_{0}$, drop the index ``$0$'' so that its number density will be simply called $n$, and define $a=1,\dots,q$. Then one can rewrite the second law as
\be
T d \sigma &=&d \left(\frac{\rho}{n}\right)+p \left(\frac{1}{n}\right)-\mth^{a} d \left(\frac{n_{a}}{n}\right)\\
&=&\frac{1}{n} \left[d\rho- \left(\frac{\rho+p}{n} \right) dn\right]-\mth^{a}  d \left(\frac{n_{a}}{n}\right)\,,
\ee
or 
\be\label{5}
n\partial_{\mu}  \sigma&=&\frac{1}{T} \left[\partial_{\mu}\rho- \left(\frac{\rho+p}{n} \right) \partial_{\mu}n-\mth^{a} n\,\partial_{\mu} \left(\frac{n_{a}}{n}\right)\right]\,.
\ee


\subsection*{Leading viscous coefficients}

It is convenient to we work in flat space and covariantize the final result to get the general relativistic expression. Let us consider the conservation laws
\be
\partial_{\mu}T^{\mu\nu}=0= \partial_{\mu}N^{\mu}_{A}\,,
\ee
for every $A=0,\dots,q$ and define as usual
\be
T^{\mu\nu}&\equiv&\left(\rho+p\right)u^{\mu} u^{\nu}+\eta^{\mu\nu}p+\Delta T^{\mu\nu}\,,\\
N_{A}^{\mu}&\equiv& n_{A} u^{\mu}+\Delta N_{A}^{\mu}\,,
\ee
where $u^{\mu}u_{\mu}=-1$ and we work in flat space. As discussed in \ref{s:Bin}, $\rho\equiv -u_{\mu}u_{\nu} T^{\mu\nu}$ and $n\equiv- u_{\mu}N^{\mu}$ and we can choose $u^{i}$ such that $u_{\mu}\Delta T^{\mu\nu}=0$ \cite{LL}. By multiplying the conservation of $T^{\mu\nu}$ by $u_{\nu}$ and using the conservation of $N_{0}$ (remember we will be dropping the label ``$0$''), one obtains
\be
u^{\mu} \left[\partial_{\mu}\rho- \left(\frac{\rho+p}{n} \right) \partial_{\mu}n\right]=-\Delta T^{\mu\nu}\partial_{\nu} u_{\mu}+ \frac{\rho+p}{n}\partial_{\mu} \Delta N^{\mu}\,,
\ee
where we used $u_{\mu}\partial_{\nu}\Delta T^{\mu\nu}=-\Delta T^{\mu\nu}\partial_{\nu} u_{\mu}$. Using this in \eqref{5} one finds
\be\label{here}
nu^{\mu}\partial_{\mu}\sigma=-T^{-1} \Delta T^{\mu \nu} \partial_{\mu} u_{\nu}+\frac{\rho+p}{T n} \partial_{\mu} \Delta N^{\mu}-\frac{\mth^{a} n}{T} u^{\mu}\partial_{\mu} \frac{n_{a}}{n}\,.
\ee
We would like to rewrite this expression as
\be\label{car}
\partial_{\mu} s^{\mu}= {\rm positive \; terms}\,,
\ee
where $s^{\mu}$ is the entropy density current (which does not make reference to any conserved charge). In this way, the entropy can only grow for an arbitrary flow. The difficulty is finding the right velocity to define the current $s^{\mu}$. We adopt the following strategy. Write some ansatz for $s^{\mu}$ and solve for the free parameters in the ansatz such that on the right hand side of \eqref{car} depends on $\Delta T$ and $\Delta N_{A}$ but not on their derivatives. Once this is done, we ask that these terms form perfect squares. This determine $\Delta T$ and $\Delta N_{A}$. 

We start with the ansatz
\be\label{ans}
s^{\mu}= s u^{\mu}+f_{A} \Delta N^{\mu}_{A}\,,
\ee
for some arbitrary functions $f_{A}$ and the sum over $A$ is implicit. We could not have added any other term, since $u_{\mu}\Delta T^{\mu\nu}$ vanishes because of our choice of $u_{\mu}$, and using derivatives gives subleading terms in the hydrodynamic expansion. Physically this ansatz say that the entropy density flow with the velocity of energy transport $u^{\mu}$ plus some corrections accounting for the difference between $u^{\mu}$ and the velocity of particle transport. Plugging \eqref{ans} into \eqref{here} one finds
\be
\partial_{\mu}s^{\mu}&=&-T^{-1} \Delta T^{\mu \nu} \partial_{\mu} u_{\nu}+\Delta N^{\mu}_{A}\partial_{\mu}f_{A}\label{smu}\\
&&\quad+\partial_{\mu}\Delta N^{\mu} \left(\frac{n_{A}\mth^{A}}{n T}+\frac{f_{A} n_{A}}{n}\right)-n u^{\mu} \left(f_{a}+\frac{\mth^{a}}{T}\right)\partial_{\mu}\left(\frac{n_{a}}{n}\right)\,,\nonumber
\ee
where we remind the reader that $n=n_{A=0}$. Choosing $f_{A}=-\mth^{A}/T$ gets rid of the second line and with it of all derivates acting on $\Delta T$ and $\Delta N$. So the final answer is
\be
s^{\mu}&=& s u^{\mu}- \frac{\mth^{A}}{T} \Delta N^{\mu}_{A}\,,\\
\partial_{\mu}s^{\mu}&=&-T^{-1} \Delta T^{\mu \nu} \partial_{\mu} u_{\nu}-\Delta N^{\mu}_{A}\partial_{\mu}  \frac{\mth^{A}}{T}.\label{last}
\ee
The first term on the right hand side of \eqref{last} captures bulk and shear viscosity corrections to $T^{\mu\nu}$. For these terms the discussion is the same as in \cite{Weinbergold}, so we do not repeat it here. Let us focus instead on $\Delta N$. A sufficient condition for the corrections to be always positive is $\Delta N^{\mu}\propto \partial_{\mu}  (\mth^{A}/T) $. From now on the discussion is the same for every $A$, so we drop the specie index. From the definition of $n$, one must have $u_{\mu}\Delta N^{\mu}=0$. To enforce this we can use the projector $H^{\mu\nu}\equiv \eta_{\mu\nu}+u^{\mu}u^{\nu}$ in $\Delta N^{\mu}\propto H^{\mu\nu}\partial_{\nu}(\mth/T)$. The constant of proportionality is a matter of convention and dimensional analysis. We use the convention of \eqref{deltaN}.

 
\section{Derivation of the heat conduction} \label{a:chi}

In this appendix we derive the formula for the heat conduction in the conservation of the number of photons $\chi_{\gamma}$ and baryons $\chi_{b}$. The latter has been known for a long time \cite{}, while the former, to the best of our knowledge, has not appeared yet in the literature.

 
\subsection*{Baryon heat conduction}

 Let us start with the Boltzmann equations for photons and baryons. Notice that we can carry on this computation in flat space and use the result in an expanding background with small perturbations. Since in the simplest models the spectral distortion is expected to be much smaller than the other perturbations, we will neglect it in this computation. The first two momenta of the two relevant Boltzmann equations are \cite{dod}
 \be
 \dot \Theta_{0}+q \Theta_{1}&=&0\,,\\
\dot \Theta_{1}-\frac{q\Theta_{0}}{3}&=&\dot \tau \left(\Theta_{1}-\frac{i v_{b}}{3}\right)\,,\\
\dot  v_{b}&=& \frac{\dot \tau}{R} \left[v_{b}+3 i \Theta_{1}\right]\,,\label{vbsol}\\
\dot  {\delta n_{b}} + i q v_{b} &=&0\,, \label{nbodB}
 \ee
where, using the notation of 
\cite{dod} $ \tau'\equiv -  t_{\gamma}^{-1}$ and we have defined the baryon velocity potential by $v_{b}^{i}=v_{b}q^{i}/|q|$. The idea is now to find the difference $\Delta u_{b}$ between the velocity appearing in the conservation of baryon number and the one appearing in the conservation of the total energy density $u$. So we write
\be
\dot  \delta = q^{2} (1+w) u\,,\quad \dot  \delta n_{b}  = q^{2}  \left(u+\Delta u_{b}\right)\,, \label{du}
\ee
where $w$ is given in \eqref{wsol},
\be
\delta\equiv \frac{\rbo \rbz+ \rgo \rgz}{\frac{4}{3}\rgz \left(1+R\right)}
\ee
 denotes perturbations to the total energy density and $\delta n_{b}$ is dimensionless. Using \eqref{nbodB} and \eqref{du} one finds
\be
q^{2} \Delta u_{b}=- \left(i q v_{b}+q^{2} u\right)\,
\ee
One can solve \eqref{vbsol} for $v_{b}$ in the tight coupling expansion. At first order the result is
\be
i q v_{b}= 3 q \Theta_{1}+\frac{3 q R \dot  \Theta_{1}}{ \dot \tau}+\mathcal{O} \left(\dot \tau^{-2}\right)\,.
\ee
Using the other Boltzmann equations and the definition of $u$ one can similarly find
\be
q^{2}u &=&-3 q \Theta_{1}-\frac{3 q R^{2} \dot  \Theta_{1}}{(1+R) \dot \tau}+\mathcal{O} \left( \dot \tau^{-2}\right)\,,\\
\dot  \Theta_{1}&=& \frac{q \Theta_{0}}{3(1+R)}-\frac{q R^{2}\dot \Theta_{0}}{3 (1+R)^{2} \dot \tau}\,.
\ee
Hence finally 
\be
\Delta u_{b}= -\frac{R}{(1+R)^{2} \dot \tau} \Theta_{0}=\frac{R}{(1+R)^{2}} t_{\gamma} \Theta_{0}\,.
\ee
One can recast this using the notation of \cite{Weinberg}, as in \eqref{deltaN}, provided that
\be
\chi_{b} T = \frac{4}{3} \rgz t_{\gamma}\,,
\ee
which agrees with \cite{Weinbergold}.

 
\subsection*{Photon heat conduction}

In order to compute the heat conduction $\chi_{\gamma}$ appearing in the conservation of photon number, we could use the same strategy as above. On the other hand, there is a shortcut that leads to the same result.  Let us start by writing the transport of energy density in two different but equivalent ways
\be
\left(\rho +p\right) u=\left(\rho_{\gamma} +p_{\gamma}\right) \left(u+\Delta u_{\gamma}\right)+\left(\rho_{b} +p_{b}\right) \left(u+\Delta u_{b}\right) \,.
\ee
By using the results of the previous section we can then find
\be
\Delta u_{\gamma}= - R \Delta u_{b} =\frac{R^{2}}{(1+R)^{2} \dot \tau} \Theta_{0}\,.
\ee
During radiation domination, when $R\ll1$, the velocity of energy transport is closer to the velocity of photon rather than baryon number transport. This is intuitive since the photons carry most of the energy.

Again one can recast this using the notation of \cite{Weinberg}, as in \eqref{deltaN}. The resulting heat conduction appearing in the conservation of photon number, caused by the presence of baryons, is
\be
\chi_{\gamma}T&=&\frac{4}{3} \rgz t_{\gamma}\,\frac{2\pi^{4}}{45\zeta(3)}\frac{R^{2}}{\bar\mu}\,,\\
&\simeq& \frac{4}{3} \rgz t_{\gamma}\, 3.6 \times \frac{R^{2}}{\bar\mu}
\ee
where the apparent divergence as $\bar \mu\rightarrow 0$ is fictitious since $\bar \mu$ cancels in the expression for $\Delta N^{\mu}$ or $\Delta u_{\gamma}$.


\section{The energy of a wave from fluid dynamics}\label{a:cs}

In this appendix we derive a relativistic formula for the energy density of a pressure wave in a perfect fluid. For simplicity we work in flat, unperturbed space. To begin we have to give a precise definition of the energy of wave up to second order in perturbations. For a perfect fluid we define
\be
E_{w}&=&\ex{\Delta T^{00}}_{p}\,,\nonumber \\
\Delta T^{00}&\equiv& \ex{T^{00} \left[\ex{\rho}=\rho_{av},u^{i}\right] }-\ex{T^{00} \left[\ex{\rho}=\rho_{av},u^{i}=0\right] }\label{Edef}\\
&=&\ex{T^{00} \left[\ex{\rho}=\rho_{av},u^{i}\right] }-\rho_{av}\,,
\ee
for some average energy density $\rho_{av}$. Here we have introduced the average over a wave period $\ex{}_{p}$. In words, the energy of a wave is the energy of the fluid with the wave ($u^{i}\neq 0$) minus the energy of the fluid without the wave ($u^{i}=0$) with the same average rest-frame energy density $\rho_{av}$. Expanding in small perturbations around a homogeneous and isotropic solution (potentially time dependent) we define
\be
\rho&=&\bar \rho\left(1+\ru+\rt\right)= \bar \rho+\delta \rho+\delta \rho_{(2)}\,,\\
\quad u_{i}&=&0+\uu+\ut\,,
\ee
so that at second order
\be\label{fin}
T^{00}= \rz \left(1+\ru+\rt \right)+(\rz+\bar p)u^{i}u_{i} \,.
\ee

Using the definition \eqref{Edef} one finds
\be\label{finu}
E_{w}\equiv \ex{\Delta T^{00}}_{p}= \ex{(\rz+\bar p)u^{i}u_{i}}_{p}\,.  
\ee
In order to rewrite this expression in terms of density perturbations, let us consider the dynamics. There are four equations $T^{\mu\nu}_{,\nu}=0$ for five variables $\rho,\, u_{i}$ and $p$. An equation of state will be needed to close the system. For a perfect fluid \eqref{emt}, energy and momentum conservation can be written as
\be
\partial_{t} \left[(\rho+p) (1+|u|^{2})-p\right]+\partial_{i} \left[(\rho+p)  u^{i} \sqrt{1+|u|^{2}}\right]&=& 0 \label{0}\,,\\
\partial_{t} \left[(\rho+p) u_{i} \sqrt{1+|u|^{2}}\right]+\partial_{j} \left[(\rho+p) u^{j} u_{i}+\delta_{ij}p\right]&=& 0\label{i}\,,
\ee
where $|u|\equiv u^{i}u_{i}$. We solve these non-linear equations in perturbation theory around a homogeneous and constant background. For this purpose it is convenient to take the time derivative of \ref{0} and solve for $\partial_{t}\partial_{i} \left[\rho u_{i} \sqrt{1+|u|^{2}}\right]$. Then one substitute this into the divergence of \ref{i}. At linear order and going to Fourier space, the result is
\be
\ddot{ \delta \rho}+q^{2} \delta p =0\,.\label{wav}
\ee
From \eqref{i} focusing on scalar degrees of freedom one finds
\be\label{wavu}
(\rz+\bar p) q^{2} u=\dot {\delta \rho}\,.
\ee

 
\subsection*{Barotropic fluid}

If we assume that the fluid is barotropic, then the equation of state takes the form $p=p(\rho)$. This in particular means that 
\be
\delta p=  \frac{\partial p}{\partial \rho} \delta \rho\equiv c_{s}^{2} \delta \rho\,,\label{baro}
\ee
where we used the definition of the speed of sound \eqref{wsol}. Notice that if the fluid had not been barotripic there would have been additional terms not dependent on $c_{s}^{2}$. The photon-baryon-electron plasma is \textit{not} in general barotropic, since (neglecting baryon temperature) $\rho=\rho_{\gamma} (4/3) (1+R)$ and $p=p_{\gamma}=\rho_{\gamma}/3$. On the other hand, it can be approximated as a barotropic fluid in at least two case. At very early times, neglecting terms of order the baryon loading $R$. During an adiabatic evolution, since then the system moves along a one dimensional adiabat, which can be parameterized by $\rho$. Hence the following formulae will be a good description of the photon-baryon-electron plasma in these limits.

Substituting \eqref{baro} into \eqref{wav} one finds oscillatoric solutions, which in the WKB approximation take the form
\be
\delta \rho=A \cos \left(\int qc_{s} dt'+{\rm phase}\right)\,.
\ee
Hence, always at leading order in the WKB expansion (i.e.~we neglect the possible time dependence of the amplitude $A$)
\be
\ex{\dot{\delta \rho}^{2} }_{p}=q^{2}c_{s}^{2}\ex{\delta \rho^{2} }_{p}\,.
\ee
Using \eqref{wavu} and \eqref{Edef} we find
\be
E_{w}&=&\ex{ (\rz+\bar p)u^{i}u_{i}}_{p}=\frac{c_{s}^{2}}{\rz+\bar p} \ex{\delta \rho^{2}}_{p} \label{rfoq}\\
&=&\rz \frac{c_{s}^{2}}{1+w} \ex{\delta^{2}}_{p}\,,\label{Ewfin}
\ee
where in the last line we used $w=\bar p/\rz$. This result agrees with the non-relativistic formula for $\bar p\ll \rz$. For radiation $c_{s}^{2}=w=1/3$, and hence 
\be\label{13}
E_{w}=(3/4) \rz c_{s}^{2}\ex{\ru^{2}}=\frac{1}{4}\rz \ex{\ru^{2}}\,.
\ee


\section{The energy of a wave from kinetic theory}\label{a:kin}

According to \cite{dod}, unlike other folks physicists have always one thing in mind: the Boltzmann equation. Let us therefore verify the hydrodynamic result \eqref{13} using kinetic theory. We start introducing our conventions and move on in the next subsection to discuss the formula for the energy of the wave. The energy-momentum tensor in an FLRW universe for a gas of bosons is
\be
T^{\mu\nu}(\vec x,t)=\frac{1}{a^{3}}\int \frac{d^{3}p}{(2\pi)^{3}} \frac{p^{\mu}p^{\nu}}{p^{0}} \frac{1}{e^{p^{0}/kT(\vec x,t,\hat p)}-1}\,,
\ee
where $p^{0}p^{0}=p_{i}p_{i}a^{-2}\equiv p^{2} a^{-2}$ and $\hat p_{i}\equiv p_{i}/p$. It is common to denote small temperature inhomogeneities by
\be
T(\vec x,t,\hat p)\equiv \bar T \left[1+\Theta(\vec x,t,\hat p) \right]\,,
\ee
where $\bar T$ is a dimensionful constant and for the moment we do not allow for distortion, i.e.~it does not dependend on $p$, but just on $\hat p$. From now on we omit to write the time dependence. We can decompose the Fourier transform 
\be
\Theta(\vec k,\hat p)\equiv \int d^{3}x \,e^{-i k_{i} x^{i}}\,\Theta(\vec x,\hat p)\,,
\ee
in multipoles 
\be
\Theta(\vec k,\hat p)&=&\Theta(\vec k,\mmu)=\sum_{l=0}^{\infty} \left(2l+1\right) \left(-i\right)^{l}\,P_{l}(\mmu) \Theta_{l}(\vec k)\,,\\
\Theta_{l}(\vec k)&=&i^{l}\int_{-1}^{1} \frac{d\mmu}{2} \Theta(\vec k,\mmu) P_{l}(\mmu)=i^{l}\int \frac{d\Omega_{\hat p}}{4\pi} \Theta(\vec k,\hat p) P_{l}(\cos \left(\theta_{\hat p}\right))\,,
\ee
where $\mmu\equiv k_{i}p_{i}/(kp)=\cos \left(\theta_{kp}\right)$ and the Legendre polynomials satisfy
\be
\int_{-1}^{1} \frac{d\mmu}{2} P_{l}(\mmu) P_{l'}(\mmu)=\frac{\delta_{ll'}}{2l+1}\,.
\ee
In the following we will also use
\be
\Theta_{l}(\vec x)&\equiv& \int \frac{d^{3}k}{(2\pi)^{3}} \,e^{i k_{i} x^{i}}\,\Theta_{l}(\vec k)\,\\
&=&i^{l}\int \frac{d\Omega_{\hat p}}{4\pi} \Theta(\vec x,\hat p) P_{l}(\cos \left(\theta_{\hat p}\right))\,.
\ee
With these definitions and conventions one finds
\be
T^{00}(\vec x)=\rz \int \frac{d^{2}\Omega_{\hat p}}{4\pi} \left[1+4 \Theta(\vec x,\hat p)+6 \Theta^{2}(\vec x,\hat p)\right]\,,
\ee
where $\rz\equiv \frac{\pi^{2}}{15}\left(kT_{r}\right)^{4}$. Decomposing in multipoles and using the statistical homogeneity of the perturbations one obtains
\be\label{100}
\ex{T^{00}}=\rz  \left[1+ 4\ex{\Theta_{0}(\vec x)}+6 \int \frac{d^{3}k}{(2\pi)^{3}}\sum_{l=0}^{\infty} (2l+1)\ex{|\Theta_{l}(\vec k)|^{2}}'\right]\,,
\ee
where 
\be
(2\pi)^{3}\delta^{3} \left(\vec k +\vec k'\right)\,\ex{|\Theta_{l}(\vec k)|^{2}}'\equiv \ex{\Theta_{l}(\vec k)\Theta_{l}(\vec k')}\,,
\ee
and we used the reality of $\tzr(\vec x)$ to substitute $\tzr(-\vec k)=\tzr^{\ast} (\vec k)$.


\subsection*{The energy of a wave}

In the tight coupling regime at every point we can go in a boosted reference frame which we denote with the label ``$r$'' (since this is the frame in which the baryons are at rest) in which
\be\label{rest}
T(\vec x,t,\hat p)\equiv T_{r} \left[1+\tzr(\vec x) \right]\,,
\ee
i.e.~there is only a monopole. Since we assume statistical homogeneity $\ex{\tzr(\vec x)}$ does not depend on $\vec x$ and can hence be absorbed into $T_{r}$. So without lost of generalities we impose $\ex{\tzr(\vec x)}=0$. $T^{00}(\vec x)$ in this comoving frame is $\rho(\vec x)$. For small inhomogeneities we have
\be
T^{00}_{r}(\vec x)\equiv\rho(\vec x)= \rz \left[1+4\tzr(\vec x)+6\tzr^{2}(\vec x)+\mathcal{O} \left(\Theta^{3}\right)\right]\,,
\ee
and therefore
\be\label{fin2}
\ex{\rho(\vec x)}&=& \rz \left[1+6\ex{\tzr^{2}(\vec x)}+\mathcal{O} \left(\Theta^{3}\right)\right]\\
&=& \rz \left[1+6 \int \frac{d^{3}k}{(2\pi)^{3}}\ex{|\tzr(\vec k)|^{2}}'+\mathcal{O} \left(\Theta^{3}\right)\right]
\ee
Notice that $\rho$ is defined as the energy density measured by an observer that sees the fluid locally at rest. Let us now review how temperature changes as we go to a boosted reference frame. If we make a boost with velocity $\vec v\equiv v_{i}$, say from reference frame $S$ to $S'$, the directional temperature changes from $T(\hat p) $ to $T(\hat p')$ according to (see e.g.~\cite{Kosowsky:2010jm})
\be\label{Ttrans}
T(\hat p)=T'(\hat p') \frac{1+\vec v\cdot\hat p'}{\sqrt{1-v^{2}}}\,,
\ee
with
\be
\hat v\cdot\hat p=\frac{\hat v\cdot\hat p'+v }{1+ \vec v\cdot\hat p'}\,.
\ee
So going from the rest frame \eqref{rest} to a boosted frame, one finds up to quadratic order in velocity
\be
\Theta(\vec x, \hat p)=\tzr(\vec x)+\vec v\cdot\hat p+ \left(\vec v\cdot\hat p\right)^{2}-\frac{v^{2}}{2}+\mathcal{O} \left(v^{3}\right)\,.
\ee
This tells us that the boost creates both a dipole and a quadrupole and modifies the monopole. More specifically, let us focus on an irrotational velocity field $v_{i}(\vec x)=\partial_{i}u(\vec x)$, so that $v_{i}(\vec k)=iku(\vec k)$. We find
\be
\Theta_{0}(\vec x)&=&\int \frac{d\Omega_{\hat p}}{4\pi} \Theta(\vec x,\hat p)=\int \frac{d^{3}k}{(2\pi)^{3}} \,e^{i k_{i} x^{i}}\,\Theta_{0}(\vec k)=\tzr(\vec x)-\frac{1}{6}v^{2}(\vec x)\,,\\
\Theta_{1}(\vec k)&=&\frac{1}{3} k\,u(\vec k)\,.
\ee
If we substitute this result into \eqref{100}, using $\ex{\tzr(\vec x)}=0$, we find
\be\label{fin1}
\ex{T^{00}}=\rz  \left\{1+ \int \frac{d^{3}k}{(2\pi)^{3}} \left[6 \ex{|\Theta_{0}(\vec k)|^{2}}'+ \left(\frac{6}{3}-\frac{4}{6}\right) k^{2}\ex{|u(\vec k)|^{2}}'\right]\right\}\,.
\ee
Using the definition of appendix \ref{a:cs} and \eqref{fin1} and \eqref{fin2} we find
\be
E_{w}&=&\rz  \int \frac{d^{3}k}{(2\pi)^{3}} \frac{4}{3} k^{2}\ex{|u(\vec k)|^{2}}'\\
&=&\rz  \frac{4}{3} \ex{u_{i}(\vec x) u^{i}(\vec x)}\,,
\ee
which, for radiation, agrees with \eqref{finu}.


\section{Evolution of distortion in the cosmological frame}\label{a:cf}

In this appendix we re-derive an equation for the time evolution of $\mu$ in the cosmological reference frame, in which the fluid as a whole does not move.  A parallel derivation using rest-frame quantities was given in subsection \ref{ss:sum}. The ensemble-averaged conservation of the energy-momentum tensor at second order is given in \eqref{fs}
\be
\ex{\partial_{t}T^{00}_{(2)}+3HT^{00}_{(2)}+a^{2}H T^{ii}_{(2)} }=0\,.
\ee
With some insight we have neglected the viscous term $\Delta T^{ii}_{(2)}$, which is subleading well inside the horizon. Approximating the plasma as a gas of photons, we can use $T^{ii}=a^{-2}T^{00}$. The conservation of photon number gives
\be
\ex{\partial_{t} N^{0}+3H N^{0}}=0\,.
\ee
Let us expand in perturbation according to
\be
T^{00}&=&\bar T^{00} \left(1+\delta T^{00}_{(1)}+\delta T^{00}_{(2)}\right)\,,\\
N^{0}&=&\bar N^{0} \left(1+\delta  N^{0}_{(1)}+\delta N^{0}_{(2)}\right)\,.
\ee
Then at second order we find the simple ensemble-averaged equations
\be\label{sim}
\partial_{t}\ex{\delta N^{0}_{(2)}}=\partial_{t}\ex{\delta T^{00}_{(2)}}=0\,.
\ee
Let us now use the expressions from kinetic theory
\be\label{ex1}
T^{00} (\vec x)&=& (kT_{0})^{4} \frac{\pi^{2}}{15} \left[1+4 \tcu+4\tct-\frac{90 \zeta(3)}{\pi^{4}}\mu+6\int \frac{d\Omega_{\hat p}}{4\pi} \tcu^{2}\right]\,,\\
N^{0}(\vec x)&=& (kT_{0})^{3} \frac{2\zeta(3)}{\pi^{2}} \left[1+3 \tcu+3\tct-\frac{\pi^{2}}{6\zeta(3)}\mu+3 \int \frac{d\Omega_{\hat p}}{4\pi} \tcu^{2}\right]\,.\label{ex2}
\ee
where the temperature in the cosmological frame is 
\be
T(\vec x, \hat p)= T_{0} \left[1+\tcu(\vec x, \hat p)+\Theta_{(2)} (\vec x, \hat p)\right]\,,
\ee
and we have used the notation
\be
\tct(\vec x)\equiv \int \frac{d\Omega_{\hat p}}{4\pi} \Theta_{(2)}(\vec x, \hat p)
\ee
Substituting \eqref{ex1} and \eqref{ex2} into \eqref{sim}, one finds
\be
\partial_{t}\ex{\mu}&=&-  \frac{9 \pi^{4}\zeta(3)}{2[\pi^{6}-405 \zeta(3)^{2}]}\times 2\partial_{t}\int \frac{d\Omega_{\hat p}}{4\pi} \ex{\tcu^{2}}\label{check}\\
&\simeq & -1.40\times2\, \partial_{t}\int \frac{d^{3}k}{(2\pi)^{3}} \left[\ex{|\Theta_{(1),0}|^{2}}+3\ex{|\Theta_{(1),1}|^{2}}\right]\\
&=&-1.40\times \frac{1}{4}\partial_{t} \ex{\rgo^{2}}\,,
\ee
where in the last line we used the wave solution to relate first order monopole and dipole $\Theta_{(1),1} \sqrt{3}=\Theta_{(1),0}=\rgo/4$ (see e.g.~\cite{Ma}). This result agrees with the estimate
\be\label{ag}
 \partial_{t}\ex{ \mu}&=& -1.40\frac{\delta E}{E}\simeq -1.40 \frac{c_{s}^{2}}{1+w}\partial_{t} \ex{\rgo^{2}}\,,
\ee
for $c_{s}^{2}=w=1/3$. The time evolution of the temperature is
\be
\partial_{t}\ex{\tct}&=& -\frac{3}{4}\,\frac{\pi^{6}-270 \zeta(3)^{2}}{\pi^{6}-405 \zeta(3)^{2}}\times 2\,\partial_{t}\int \frac{d\Omega_{\hat p}}{4\pi} \ex{\tcu^{2}} \label{check3}\\
&\simeq & -1.14\times2\, \partial_{t}\int \frac{d^{3}k}{(2\pi)^{3}} \left[\ex{|\Theta_{(1),0}|^{2}}+3\ex{|\Theta_{(1),1}|^{2}}\right]\\
&=&-1.14\times \frac{1}{4}\partial_{t} \ex{\rgo^{2}}\,.
\ee
The above derivation, performed in the cosmological frame, agrees with the one presented in subsection \ref{ss:sum} performed in the rest frame. First, $\mu$ does not depend on the reference frame, and in fact \eqref{ag} is identical to \eqref{qf}. Second, given that the temperature transforms as in \eqref{Ttrans}, we expect
\be
\ex {\tct- \trt}=- \frac{1}{6}\ex{v^{2}}=- \frac{1}{8}\times \frac{1}{4}\partial_{t} \ex{\ru^{2}}\,.
\ee
From 
\be
\partial_{t}\ex{ \trt}&=& \frac{ \left(405 \zeta(3)-2\pi^{6}\right)  \partial_{t} \ex{u^{i}u_{i} }+ 9 \left(270\zeta(3)^{2}-\pi^{6}\right)\partial_{t} \ex{\tru^{2} }}{6 \left[\pi^{6}-405 \zeta(3)^{2}\right]}\nonumber\\
&\simeq &- 0.59 \partial_{t} \ex{u^{i}u_{i} }- 2.3\partial_{t} \ex{\tru^{2} } \,.
\ee
 and \eqref{check3} one can check that this is indeed the case.

 
\section{The $\mu$ power spectrum}\label{a:ps}

In this appendix we give a derivation of the three-dimensional power spectrum of $\mu$ at the end of the $\mu$-era. This can be used together with the transfer function, which we discuss in subsection \ref{Afs}, to obtain the late time $l$-space power spectrum. As explained in subsection \ref{Afs}, only the power spectrum $\ex{\mu(\vec q)\mu(\vec q)}$ in the limit $q\rightarrow 0$ affects observations on angular scales relevant for observation. Hence in the solution \eqref{mufin}, we can drop the last term since it is a total derivative. Then we can easily integrate over time and find
\be
\mu_{0}(\vec x,t_{f})= \frac{2}{3\Ao-4\At}  \left[u_{i}u^{i}+3\tru^{2}\right]^{f}_{i}\,,
\ee
where $t_{f}=t(z_{f})$ and $ \left[\cdot\right]^{f}_{i}$ indicates the difference of its argument between the beginning of the $\mu$-era at $z_{i}\simeq 2\times 10^{6}$ and the end $z_{f}\simeq 5\times 10^{4}$. We now use $\tru=\rgo/4$, $4qu/a=3\rgod$ and the explicit expression for $\rgo$ in \eqref{again2}. Neglecting terms of order $R$, the Fourier transform of $\mu$ is then
\be
\mu_{0}(\vec q,t_{f})&=& \frac{2}{3\Ao-4\At}\int \frac{d^{3}k}{(2\pi)^{3}} 9 c_{s}^{2} \R^{0}(\vec k)\R^{0}(\vec q-\vec k) \left(1-0.268 R_{\nu}\right)^{2} \\
&&\quad   \left[\sin \left(\int^{t_{f}} \frac{k}{a}c_{s}\,dt'\right)\sin \left(\int^{t_{f}} \frac{|\vec q-\vec k|}{a}c_{s}\,dt'\right) e^{- \left(k^{2}+|\vec q-\vec k|^{2}\right)q_{D}^{-2}}+\right. \\
&&\quad \left.+\cos \left(\int^{t_{f}} \frac{k}{a}c_{s}\,dt'\right)\cos \left(\int^{t_{f}} \frac{|\vec q-\vec k|}{a}c_{s}\right)e^{- \left(k^{2}+|\vec q-\vec k|^{2}\right)q_{D}^{-2}}\right]^{f}_{i}\nonumber\,.
\ee
We can use this expression to compute the momentum space two-point correlation function in the limit $q\rightarrow 0$
\be
\ex{\mu_{0}(\vec q,t_{f})\mu_{0}(\vec q',t_{f})}=(2\pi)^{3}\delta^{3} \left(\vec q+\vec q'\right)  \frac{2\left(1-0.268 R_{\nu}\right)^{2}}{3 \left(3\Ao-4\At\right)} \int \frac{d^{3}k}{4\pi} \frac{2\pi^{2}\Delta_{\R}^{4}(k)}{k^{6}}  \left\{ \left[ e^{-2k^{2}/q_{D}^{2}}\right]^{f}_{i}\right\}^{2}\,.\nonumber
\ee
For a primordial power spectrum close to scale invariance, the integral is manifestly supported on the largest values of $k$. For an exactly scale invariant primordial power spectrum of amplitude $\Delta_{\R}^{2}(k_{p})$, this reduces to
\be
P_{\mu_{0}}(q\rightarrow 0,\tau_{f})&=& \frac{2\left(1-0.268 R_{\nu}\right)^{2}}{3 \left(3\Ao-4\At\right)} \int \frac{d^{3}k}{4\pi} \frac{2\pi^{2}\Delta_{\R}^{4}(k)}{k^{6}}  \left\{ \left[ e^{-2k^{2}/q_{D}^{2}}\right]^{f}_{i}\right\}^{2} \label{Pmu}\\
&\simeq&2\times 10^{-14}\,  \frac{2\left(1-0.268 R_{\nu}\right)^{2}}{3 \left(3\Ao-4\At\right)} 2\pi^{2}\Delta_{\R}^{4}(k_{p})\\
&\simeq&\frac{4}{q_{D}(z_{f})^{3}}\, \frac{2\left(1-0.268 R_{\nu}\right)^{2}}{3 \left(3\Ao-4\At\right)} 2\pi^{2}\Delta_{\R}^{4}(k_{p})\,,
\ee
which we simply call $P_{\mu_{0}}(\tau_{f})$.


\end{document}